\documentclass[journal]{IEEEtran}
\IEEEoverridecommandlockouts

\usepackage[linesnumbered]{algorithm2e}
\usepackage{enumitem}
\usepackage{color}
\usepackage{theorem}
\usepackage{times,amsmath,epsfig}
\usepackage{amssymb}
\usepackage{cite}
\usepackage{multirow}
\usepackage{comment}

\usepackage{hyperref}
\usepackage{algorithmic}
\usepackage{rotating}
\usepackage{cleveref, subcaption}

\usepackage[utf8]{inputenc}
\usepackage{color}
\usepackage{theorem}
\usepackage{times,amsmath,epsfig}
\usepackage{multirow}
\usepackage{amsfonts}
\usepackage{spalign}
\usepackage[linesnumbered]{algorithm2e}

\usepackage{graphicx}
\usepackage{subcaption}

\usepackage[font=small]{caption}

\setlist[itemize]{leftmargin=*}

\input{mysymbol.sty}

\newenvironment{myproof}
  {\noindent \textbf{Proof:} }
  {\hfill$\square$}

\title{Learning the Topology of a Simplicial Complex Using Noisy Simplicial Signals}
\author{Andrei Buciulea, ~\IEEEmembership{Member,~IEEE}, Elvin Isufi,~\IEEEmembership{Senior Member,~IEEE}, Geert Leus,~\IEEEmembership{Fellow,~IEEE} and Antonio G. Marques,~\IEEEmembership{Senior Member,~IEEE} \vspace{-0.8cm}\thanks{Work supported by the EU H2020 Grant Tailor (No 952215, agreements 31 and 82); the Dutch Grant GraSPA (No 19497) financed by the Netherlands Organization for Scientific Research (NWO); the  TU Delft AI Initiative; the SURE-AI center grant 357482 financed by the Research Council of Norway, the Spanish  (MCIN/AEI/10.13039/501100011033) grant PID2022-136887NB-I00; and by the Autonomous Community of 
Madrid within the ELLIS Unit Madrid framework, and the grants URJC/CAM F1180 (Linea A. CP2301) and TEC-2024/COM-89. Preliminary results were reported in the conference paper \cite{buciulea2024learningsam}.~~A. Buciulea \& A.G. Marques are with King Juan Carlos University, Madrid, Spain (e-mail:\{andrei.buciulea, antonio.garcia.marques\}@urjc.es).     
E. Isufi \& G. Leus are with the Delft University of Technology, Delft, The Netherlands (e-mail:\{e.isufi-1, g.j.t.leus\}@tudelft.nl).}}

\input{mysymbol.sty}

\def \ccalEobs {\ccalE^{\ccalO}}

\newcommand{\vvec}{{\mathrm{vec}}}

\newcommand \blue[1] {#1}

\usepackage{mathtools}
\DeclarePairedDelimiterX{\norm}[1]{\lVert}{\rVert}{#1}

\newtheorem{theorem}{Theorem}

\newtheorem{mylemma}{\bf Lemma}

\newtheorem{remark}{\bf Remark}

\begin{document}

\maketitle

\begin{abstract} 
Graphs are a fundamental tool for modeling the irregular (non-Euclidean) structure of complex data. However, they are inherently limited to representing pairwise relationships, making them inadequate for datasets exhibiting higher-order interactions. Simplicial complexes (SCs) have emerged as a promising framework for capturing such higher-order dependencies. This paper focuses on the problem of identifying the topology of an SC from signals, which serves as the foundation for SC-based processing and learning schemes. We consider a setting where we observe noisy signals (features) associated with the nodes of the SC (0-simplices) and a subset of the edges (1-simplices). We assume the observed signals are \textit{smooth} over the unknown SC topology, and that the higher-order interactions are sparse. Building on these assumptions, we formulate topology learning as a nonconvex optimization problem and propose an efficient block-coordinate descent (BCD) algorithm to solve it. A key step in our formulation is the modeling of the topology of the SC using binary edge and triangle selection vectors, combined with efficient greedy algorithms for optimizing such vectors. We establish theoretical convergence guarantees to a stationary point \blue{of a relaxed (penalized) version of the problem} and discuss computational complexity. 
Multiple numerical experiments with both synthetic and real-world datasets validate the effectiveness of our approach, highlighting the capability of SC-learning methods to uncover and model higher-order relationships in complex datasets. 
\end{abstract}

\section{Introduction}

Graphs have become a cornerstone of data science over the past decade, providing a powerful and effective way to model pairwise relationships in complex data \cite{friedman2008sparse,timme2007revealing,mateos2019connecting,shuman2013emerging}. However, many real-world phenomena involve interactions that extend beyond simple pairwise relationships. This insight has prompted growing interest in higher-order network models that capture multi-way interactions \cite{battiston2021physics}. In particular, \emph{simplicial complexes} (SCs) have emerged as an attractive framework for higher-order modeling. An SC is a collection of nodes, edges, triangles, and higher-order simplices that is closed under inclusion. That is, whenever a higher-order relationship exists (e.g., a triangle), all its lower-order facets (edges and nodes) are also included. This hierarchical structure lies between graphs and general hypergraphs. It is more structured and interpretable than an arbitrary hypergraph, which imposes no inclusion rules, yet more expressive than a simple graph \cite{giusti2016twos}. The algebraic structure of SCs enables using tools from algebraic topology and geometry, such as boundary operators and higher-order Laplacians. 
Consequently, SCs have gained popularity in data science as a principled way to incorporate higher-order relationships \cite{barbarossa2020topological,schaub2021SPOnHigherOrder,yang2022simplicial,isufi2025topological}, \blue{and to facilitate learning on SCs~\cite{wu2024sclearning,gurugubelli2024sann,yang2025hodge}.}
\blue{They have been employed to model multi-neuron firing patterns in brain networks \cite{giusti2016twos}, analyze collaboration and co-authorship networks via higher-order connectivity \cite{patania2017shape}, and improve link prediction by considering triangle “closures” in social networks \cite{benson2018simplicial}.}

While graphs and SCs provide powerful models, their utility depends critically on knowing the correct topology—that is, which edges or higher-order simplices are present. \emph{Topology inference} from data has therefore become a fundamental task. Succinctly, the key question is how to use given signal observations associated with the nodes and potential higher-order entities to infer the underlying connectivity structure. \blue{This question is approached by several works on graph and hypergraph topology inference as we detail next.}

\noindent \textbf{Graph inference:} For graph identification, a variety of methods have been developed, often by imposing structural priors on the signals. Traditional approaches construct graphs by estimating pairwise similarities, partial correlation, or the precision matrix under sparsity constraints \cite{friedman2008sparse,egilmez2017graph}. In recent years, the graph signal processing (GSP) community has proposed techniques that leverage models of graph signals to infer edges \cite{dong2019learning, mateos2019connecting}. A common assumption is that the observed signals are \emph{smooth} w.r.t. the graph, meaning that connected nodes exhibit similar signal values. \blue{Under this assumption, one can formulate  problems to find a graph Laplacian that minimizes a graph smoothness objective \cite{kalofolias2016learn,dong2016learning,chepuri2017learning,saboksayr2021online}.} Other methods take advantage of graph filtering processes \cite{thanou2017learning} or assume the observed signals are stationary on the graph \cite{segarra2017network, buciulea2025polynomial}. \blue{For a more detailed overview of graph inference methods we refer the reader to \cite{mateos2019connecting,dong2019learning,giannakis2018topology}.}

\noindent \textbf{Hypergraph inference:} In contrast, learning the topology of hypergraphs or SCs from data is far less mature. A common approach is to infer higher-order structures by examining pairwise similarities. For example, one might add a 2-simplex (triangle) connecting three nodes if all three pairwise edges among them are strong according to some metric \cite{giusti2016twos}. While straightforward, these proxy methods effectively reduce the problem to graph inference followed by a post-processing step (e.g., forming cliques or hyperedges from densely connected subgraphs). As a result, they may overlook genuinely higher-order effects that are not evident in pairwise marginals. Canonical examples of this class of methods arise in topological data analysis. The Vietoris--Rips, \v{C}ech, and $\alpha$-complexes construct SCs from pairwise distances (via clique formation, common ball intersections, or Delaunay-based geometric filtering, respectively), with the goal of recovering the underlying topological structure of the data \cite{ghrist2008barcodes,edelsbrunner2003shape}. Despite their theoretical appeal, all these constructions fundamentally rely on pairwise distances and thresholding rules. Some works extend the graph smoothness prior to hypergraphs by assuming that nodes within the same hyperedge share similar features \cite{tang2023hypergraphs} \blue{or by combining smoothness measures on pairs of nodes to hyperedges~\cite{nguyen2021learning}.} Another emerging direction employs time-series dynamics: by observing the evolution of node states, one can infer which higher-order interactions (hyperedges or simplices) are necessary to explain the dynamics \cite{delabays2025hypergraph}. \blue{Finally, our recent work \cite{buciulea2024learningicassp} proposes a Volterra autoregressive model that jointly infers graph edges and 2-simplices from nodal observations.}

Despite these advances, most existing techniques are limited to node-level observations and simplistic criteria, rarely exploiting topological priors or the inherent closure structure of SCs.
This limitation restricts their ability to leverage richer sources of information, such as signals defined on edges or higher-order simplices. As a consequence, such methods often rely on stronger assumptions---e.g., that nodal data alone encode all relevant dependencies---and fail when interactions are expressed through higher-order dynamics. Moreover, they rarely exploit \emph{topological priors}, i.e., structural constraints arising from the algebraic closure properties of SCs. For example, the fact that a triangle can only exist if its three edges are present, or that incidence relations between simplices follow consistency rules. Incorporating such priors reduces the ambiguity of the inference problem and ensures the learned complexes respect meaningful combinatorial and homological structure, which in turn enhances interpretability and downstream analysis.

\noindent\textbf{Simplicial complex inference:} In the context of SCs, recent works address the problem of inferring simplicial structures directly from data. 
\blue{For instance, \cite{barbarossa2020topological} recovers three-way interactions by leveraging edge-flow smoothness, while \cite{hoppe2024representing} maximizes the fit of the edge-flow to the curl space. In a related direction, \cite{gurugubelli2024simplicial,sardellitti2023probabilistic,marinucci2026simplicial} introduce a probabilistic model for SC inference, linking node, edge, and triangle signals. 
Further,~\cite{wang2022full} develops a maximum-likelihood framework to reconstruct SCs from binary contagion dynamics.}

Overall, these methods illustrate a growing interest in learning SC topologies from diverse data modalities, ranging from nodal signals to edge flows and temporal dynamics. \blue{However, most of these approaches rely on restrictive assumptions: some consider that the underlying graph is known in advance and that the edge signals are fully observed \cite{barbarossa2020topological,hoppe2024representing,sardellitti2023probabilistic,marinucci2026simplicial,gurugubelli2024simplicial}, while others assume only nodal information is available \cite{buciulea2024learningicassp,wang2022full}. }
\blue{These conditions do not always hold in practice, which highlights the need for robust methods capable of jointly inferring both the graph and the simplicial structures from partial node and partial edge observations.}

Lastly, it is important to recognize that inferring higher-order topologies presents significant computational and statistical challenges. The number of possible higher-order interactions grows combinatorially with the number of nodes, making brute-force approaches impractical. Efficient algorithms must therefore exploit structural properties or domain knowledge to prune the search space, or impose sparsity constraints to consider only a limited subset of candidate interactions. Statistically, inferring higher-order connectivity is inherently data intensive. Higher-order models possess many more degrees of freedom than graphs (potential hyperedges and simplices), in which a reliable estimation demands a commensurate amount of data. These challenges necessitate careful algorithmic design to render the SC topology inference problem tractable in practical scenarios.

\noindent \textbf{Contributions:} This paper focuses on inferring the structure of a second-order SC from noisy and partially observed node and edge signals. A key novelty of our approach is the ability to handle partial observations, a setting largely overlooked in prior SC learning methods that typically assume full and noise-free access to nodal or edge data. We assume full access to noisy (perturbed) node signals alongside a subset of noisy edge signals, with the objective of recovering both the missing edges and the triangle connections. To achieve this, we impose classical sparsity assumptions on the SC structure and enforce smoothness constraints in two complementary ways: (i) for node signals, smoothness is promoted by connecting nodes that exhibit similar values, consistent with common practices in graph-based learning; and (ii) for edge signals, smoothness is imposed through the simplicial structure, where filled triangles (2-simplices) are encouraged to exhibit small curl values, reflecting approximately curl-free behaviors often observed in real-world flow data. 

The resulting formulation is highly nonconvex, prompting us to introduce a suitable relaxation followed by a block-coordinate descent (BCD) algorithm. Our proposed framework yields a novel four-step algorithm, wherein edges and triangles are identified via a computationally efficient greedy scheme. We establish the optimality of each step when restricted to the corresponding block of variables and further show that the overall procedure converges to a stationary point \blue{of a penalized version of the original problem.} A preliminary version of this work was presented in \cite{buciulea2024learningsam}. This extended version considers a more general setting, jointly designs the learning of the SC and the restoration of missing signals, includes optimality proofs, provides a formal convergence analysis, and offers extensive numerical experiments.

\noindent \textbf{Outline:} Section \ref{Sec:Nota_and_Prelims} introduces notation and fundamentals of SCs. Section \ref{Sec:ProblemFormulations} defines our SC-learning problem and formulations to address it. Section \ref{S:GreedyAlgorithmAlgorithm} designs a BCD algorithm to solve the optimization problem in an efficient manner and provides theoretical convergence results. Numerical results in Section \ref{Sec:NumericalResults} (including both synthetic and real-world data) and conclusions in Section \ref{Sec:Conclusions} close the paper. 

\section{Notation preliminaries}\label{Sec:Nota_and_Prelims}

This section introduces notation, reviews the fundamentals of SCs and discusses the notion of curl-smooth edge signals.

\begin{figure}[h]
    \centering

    \begin{subfigure}[b]{0.3\textwidth}
        \includegraphics[width=\textwidth]{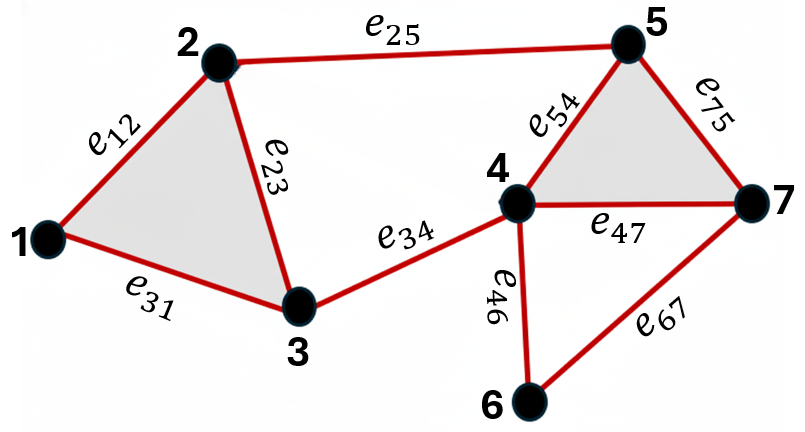}
    \end{subfigure}

    \caption{Representation of an SC with \(N = 7\) nodes (0-simplices), \(E = 10\) edges (1-simplices), and \(T = 2\) filled triangles (2-simplices), \(\{(1,2,3), (4,7,5)\}\).}

    \label{fig:SC_representation}
\end{figure}

\vspace{2mm}
\noindent \textbf{SCs and incidence matrices:}
Let $\ccalN$, $\ccalE$, and $\ccalT$ be the sets of nodes, edges, and filled triangles, respectively. We denote the number of nodes, edges and triangles as $N=|\ccalN|$, $E=|\ccalE|$, and $T=|\ccalT|$, respectively. Here, $\ccalE \subset \ccalN \times \ccalN$ comprises the active edges connecting nodes, and $\ccalT \subset \ccalN \times \ccalN \times \ccalN$ contains the active (filled) triangles formed by triplets of nodes. For a triplet $(\ccalN,\ccalE,\ccalT)$ to qualify as an SC of order~2, it must hold that whenever $(v_1,v_2,v_3)\in\ccalT$, the three edges $(v_1,v_2)$, $(v_1,v_3)$, and $(v_2,v_3)$ belong to $\ccalE$. We refer to the elements of $\ccalN$, $\ccalE$, and $\ccalT$ as simplices of order 0, 1, and 2, respectively. See Fig.~\ref{fig:SC_representation} for a graphical example of the elements of an SC. For undirected SCs, the maximum possible numbers of edges and triangles are $\bar{E}={N \choose 2}$ and $\bar{T}={N \choose 3}$, respectively.

As with graphs, the topology of an SC can be encoded in Laplacian-based matrices, which we introduce next. Let $\bbB_1 \in \reals^{N\times E}$ and $\bbB_2 \in \reals^{E\times T}$ be the incidence matrices describing the membership of nodes, edges, and triangles in the SC~\cite{schaub2021SPOnHigherOrder}:
\begin{itemize}[leftmargin=*,topsep=0pt,itemsep=0pt]
    \item Each column of $\bbB_1$ corresponds to an active undirected edge indexed by an unordered pair $\{i,j\}$ with $i<j$. An arbitrary but fixed orientation is assigned to each edge (e.g., $i\rightarrow j$), and the corresponding column has exactly two nonzero entries, $+1$ and $-1$, at the rows associated with nodes $i$ and $j$.
    \item Each column of $\bbB_2$ represents an active (filled) triangle indexed by an unordered triple ${i,j,k}$ with $i<j<k$. A fixed cyclic orientation (e.g., $(i,j,k)$) is chosen, and the three nonzero entries in the column correspond to the triangle’s edges, taking values $\pm1$ depending on orientation agreement. 
\end{itemize}
From these oriented incidence matrices, we define
\[
    \bbL_0 \;=\; \bbB_1 \bbB_1^\top,
    \quad
    \bbL_1 \;=\; \bbB_1^\top \bbB_1 \;+\;\bbB_2 \bbB_2^\top, \quad  \bbL_2 \;=\; \bbB_2^\top \bbB_2,
\]
where $\bbL_0$ denotes the (node) combinatorial Laplacian, and $\bbL_1$ is the so-called \emph{Hodge Laplacian} composed of two terms: the lower Laplacian and the upper Laplacian. The lower Laplacian $\bbB_1^\top \bbB_1$ captures the interactions between edges that share a common node, thus encoding how edges are connected through vertices. The upper Laplacian $\bbB_2 \bbB_2^\top$ captures the interactions between edges that belong to a common triangle, describing how edges are coupled through higher-order structures of the SC. Finally, $\bbL_2 = \bbB_2^\top \bbB_2$ captures the relationships between triangles that share an edge, characterizing the connectivity and orientation consistency among 2-simplices. Together, these Laplacians enable the analysis of signals defined not only on nodes but also on edges and higher-dimensional simplices within the SC.

Valid incidence matrices satisfy the \emph{simplicial complex closure} (SCC) property, meaning that a triangle can be filled only if its three associated edges exist. This also implies
\begin{equation}\label{E:simp_clos_constr}
	\bbB_1\,\bbB_2 \;=\;\bb0.
\end{equation}
From this relationship, it follows that the range spaces of $\bbB_1^\top$ and $\bbB_2$ are orthogonal, implying that any edge-space vector in $\reals^E$ admits the orthogonal decomposition $\reals^E \;=\; \mathrm{span}(\bbB_1^\top)\;\oplus\;\mathrm{span}(\bbB_2)\;\oplus\;\mathrm{kernel}(\bbL_1)$.

\vspace{2mm}
\noindent \textbf{SC signals:} 
We consider signals associated with the nodes and edges of the SC. Let \(\bbX_0 \;=\; [\,\bbx_1^0,\dots,\bbx_{P_0}^0] \;\in\;\mathbb{R}^{N\times P_0}\) represent $P_0$ nodal observations per node. In practice, a noisy version of this matrix is observed, which is denoted by \(\bbX_0^\ccalO = [\bbx_1^{\ccalO,0},...,\bbx_{P_0}^{\ccalO,0}] \in\mathbb{R}^{N\times P_0}\).
Next, assume that \(\bbX_1 = [\,\bbx_1^1,\dots,\bbx_{P_1}^1] \in\mathbb{R}^{E\times P_1}\) contains the full set of $P_1$ edge signals. However, among the $E$ active edges, let $\ccalEobs\subset \ccalE$ (of cardinality $E^{\ccalO}<E$) be the set of observed edges, and $\ccalE^{\ccalU}=\ccalE\setminus \ccalEobs$ be the set of unobserved edges (with cardinality $E^\ccalU=E-E^\ccalO$). 
Then, \(\bbX_1^\ccalO = [\bbx_1^{\ccalO,1},...,\bbx_{P_1}^{\ccalO,1}] \in\mathbb{R}^{E^\ccalO\times P_1}\) collects the observed noisy edge (flow) signals. 
If $\bbTheta \in \{0,1\}^{E^\ccalO\times{E}}$ is the edge sampling matrix (selecting rows of the $E\times E$ identity $\bbI_E$ corresponding to $\ccalEobs$), then $\bbX_1^\ccalO \!\approx \bbTheta\,\bbX_1$ (equality up to noise).

\vspace{2mm}
\noindent \textbf{Smoothness:} 
Nodal observations are considered smooth if they vary little across edges. A common measure of nodal smoothness is via the Laplacian quadratic form
\begin{equation}\label{L0smooth}
\sum_{p=1}^{P_0} \bbx_p^{0 \top} \bbL_0 \bbx_p^0 = \text{trace} (  \bbX_0^\top   \bbL_0  \bbX_0 ) =  \text{trace} (  \bbX_0 \bbX_0^\top \bbB_1\bbB_1^\top ), 
\end{equation}
which decreases as the nodal signals become smoother~\cite{shuman2013emerging,mateos2019connecting}.    
Next, the earlier expression $\reals^E=\mathrm{span}(\bbB_1^\top)\oplus\mathrm{span}(\bbB_2)\oplus\mathrm{kernel}(\bbL_1)$ allows us to decompose any edge signal $\bbx^1$ as 
$$\bbx^1 = \bbB_1^\top \tilde{\bbx}^0 + \bbB_2 \tilde{\bbx}^2 + \tilde{\bbx}^1_{\rm h},$$
The terms of this so-called Hodge decomposition can be interpreted as follows.
\begin{itemize}
\item The term $\bbB_1^\top \tilde{\bbx}^0$ is the gradient flow and is induced by the difference of the node signal $\tilde{\bbx}^0$ over the edges. Related to this we can define the divergence 
operator $\bbB_1 \bbx^1$ which measures the divergence of an edge
flow. The $i$th element corresponds to the flow passing through the $i$th node. If $\bbB_1 \bbx^1 = \bb0$ the flow is divergence-free.
\item On the other hand, $\bbB_2 \tilde{\bbx}^2$ is the curl flow and consists of the cyclic flow along the edges of all active triangles induced by the triangle signal $\tilde{\bbx}^2$. Related to this, we can define the curl operator $\bbB_2^\top \bbx^1$ which yields a triangle signal that measures the curl of an edge flow. The $i$th element corresponds to the sum of the flows of each edge forming the $i$th triangle. If $\bbB_2^\top \bbx^1 = \bb0$ the flow is curl-free.
\item The remaining term $\tilde{\bbx}^1_{\rm h}$ is called the harmonic flow. Such flows are obtained by solving the equation $\bbL_1\tilde{\bbx}^1_{\rm h} = \bbzero$. 
\end{itemize}
Real flow signals often have a small divergence or curl. 
For instance, in transportation or flow networks, traffic or fluid streams are often approximately divergence-free due to conservation laws, while in electrical circuits or consensus dynamics, edge signals exhibit low curl since cyclic imbalances are energetically costly or unstable \cite{lim2020hodge}. In this paper, we assume that flows have a small curl, since this provides useful structural information on $\bbB_2$. This assumption holds for edge signals that concentrate most of their energy in slow-varying curl components, making the curl representation approximately sparse.

Specifically, assuming that the edge signal $\bbX_1$ has a small curl (or is ``smooth'' over the triangles) means that the following measure should be small \cite{barbarossa2020topological,yang2022simplicial}: 
\begin{equation}\label{L1smooth}
     \|  \bbB_2^\top \bbX_1 \|_F^2  = \text{trace} (  \bbX_1 \bbX_1^\top \bbB_2\bbB_2^\top ).
\end{equation} 

Based on the definitions of nodal and edge smooth signals in \eqref{L0smooth} and \eqref{L1smooth}, we next describe the steps for learning SCs from nodal and edge signals, assuming that these signals are smooth over the edges and triangles, respectively.

\section{An optimization approach for learning SC from signal observations}\label{Sec:ProblemFormulations}

This section formalizes the SC-learning problem. We first introduce a preliminary optimization-based formulation grounded in incidence matrices and discuss its associated challenges. We then propose an alternative formulation that replaces the incidence matrices with edge and triangle selection vectors and introduces a novel form for the bilinear SCC constraint \eqref{E:simp_clos_constr}. This alternative formulation will be the main focus of the paper.

\subsection{Problem statement and naive formulation}

\noindent \textbf{Problem 1} \emph{Given a set of nodes $\ccalN$, a collection of noisy node-signal observations $\bbX_0^\ccalO$, and a collection of partial and noisy edge-signals $\bbX_1^\ccalO$ observed at some edges ($\ccalEobs = \ccalE \setminus \ccalE^{\ccalU}$) of an SC of order 2; find the sets $\ccalE$ and $\ccalT$ that, together with $\ccalN$, define the SC, under assumptions (AS1-AS4): \\
\noindent (AS1) The number of edges is small;\\ 
\noindent (AS2)  The number of filled triangles is small;\\ 
\noindent (AS3) The node-observations $\bbX_0$ are smooth on the graph; and \\
\noindent (AS4) The edge-observations $\bbX_1$ have a low curl (or are ``smooth'' over the triangles).
}

(AS1) is standard and just requires the graph Laplacian matrix $\bbL_0$ to be sparse. (AS2) is the natural counterpart to (AS1), requiring the upper Laplacian $\bbB_2\bbB_2^\top$ to be sparse. (AS3) establishes a link between the nodal observations and the edges, which has been used before and can be related to Gaussian modelling of attractive Markov random fields \cite{dong2016learning,egilmez2017graph,mateos2019connecting}. Finally, (AS4) is not only a prudent counterpart to (AS3) but forces the flow to have a low-curl component as discussed earlier.




Incorporating assumptions (AS1–AS4), we propose the following nonlinear optimization formulation for the SC-learning problem. This formulation promotes sparsity and signal smoothness, ensures the underlying SC structure, and accounts for the observed edge labels
\begin{subequations}\label{E:SC_OPT_v0}
\begin{alignat}{2}%
	\!\!\!\!&\! \min_{\{\bbB_i\}_{i=1}^2,\{\bbX_i\}_{i=0}^1} &&  ~\|\bbB_1\bbB_1^\top\|_0 + \|\bbB_2\bbB_2^\top\|_0  \nonumber\\
	\!\!\!\!&\!\! &&+ \|\bbX_0-\bbX_0^\ccalO\|_F^2 +\|\bbTheta\bbX_1-\bbX_1^\ccalO\|_F^2     \nonumber\\	
    \!\!&\!\! &&+\tr(\bbX_0\bbX_0^\top\bbB_1\bbB_1^\top)\!+ \!\tr(\bbX_1\bbX_1^\top\bbB_2\bbB_2^\top)   \label{E:SC_OPT_v0_obj}\\
	\!\!\!\!&\! \hspace{4.5mm}\mathrm{\;\;s. \;t. } && 
	\bbB_1\in \ccalB_1 ,\;\bbB_2\in \ccalB_2, \label{E:SC_OPT_v0_c1}\\
	\!\!\!\!&\! && \bbB_1\bbB_2 = \bb0, \label{E:SC_OPT_v0_c2}\\
	\!\!\!\!&\! && [\bbB_1\bbB_1^\top]_{ij}=-1 \;\;\blue{\text{for all}} \;(i,j)\in \ccalEobs, \;\; \label{E:SC_OPT_v0_c3}
\end{alignat} 
\end{subequations}
\noindent
where $\bbTheta\in\{0,1\}^{E^{\ccalO}\times E}$ is the edge sampling matrix, and $\ccalB_1$, $\ccalB_2$ denote the sets of feasible incidence matrices. \blue{Note that the number of edges to be estimated, $E$, is unknown and depends on the optimization variable $\bbB_1$.}  The first two terms in the objective reflect the sparsity assumptions in (AS1) and (AS2), the third and fourth terms capture the noisy nodal signals and the noisy and missing edge signal observations, and the two last terms handle the smoothness requirements in (AS3) and (AS4). For readability, we have omitted the regularizer weights. \blue{Constraints~\eqref{E:SC_OPT_v0_c1}--\eqref{E:SC_OPT_v0_c2} impose necessary conditions for a valid SC. Specifically, the minimum cardinality requirements for edges and triangles (to prevent the trivial all-zero solution) are assumed to be embedded within the definitions of the feasible sets $\ccalB_1$ and $\ccalB_2$, while~\eqref{E:SC_OPT_v0_c2} enforces the algebraic necessary condition for the SCC. Finally,} constraint~\eqref{E:SC_OPT_v0_c3} incorporates the observed edges. 

Formulation~\eqref{E:SC_OPT_v0} presents three primary challenges: 1) it is nonconvex due to the $\ell_0$ norms and the multilinear terms; 2) enforcing that $\bbB_1$ and $\bbB_2$ are feasible incidence matrices is challenging by design; and 3) ensuring that the solution satisfies the SCC 
property is nontrivial. To elaborate on this last point, consider the case where the set of active links is known and $\bbB_1$ only indexes those links. Then, having $\bbB_1\bbB_2=\bb0$ implies SCC. However, if, as customary in graph learning approaches, one uses an upper bound on the number of active links and then sets some of the columns of $\bbB_1$ to zero, it can be shown that a $\bbB_2$ matrix that activates a triangle whose 3 links are zero in $\bbB_1$ still yields $\bbB_1\bbB_2=\bb0$. In other words, the constraint $\bbB_1\bbB_2=\bb0$ is necessary for 
SCC, but not sufficient. Section~\ref{S:Greedy_SCL_formulation} introduces a tractable approach to address these challenges.

\subsection{SC-learning approach via simplex selection}\label{S:Greedy_SCL_formulation}
Our two main key ideas for making the optimization in \eqref{E:SC_OPT_v0} more tractable are: K1) recast the optimization over the incidence/Laplacian matrices as an edge/triangle selection problem; and K2) enforce the SCC via a constraint.

To describe K1), consider the \emph{complete} SC $(\ccalN,\bar{\ccalE},\bar{\ccalT})$, where 
\blue{\[\bar{\ccalE}=\{(i_1,i_2):1\leq i_1 < i_2 \leq N\}\;\text{and}\]
\[\bar{\ccalT}=\{(i_1,i_2,i_3):1\leq i_1 < i_2 < i_3 \leq N\}.\]} 
Let $\bar{\bbB}_1\in\reals^{N\times \bar{E}}$ and $\bar{\bbB}_2\in\reals^{\bar{E}\times \bar{T}}$ denote the incidence matrices of this complete SC. Now consider any (non-complete) SC $(\ccalN,\ccalE,\ccalT)$ and associate two binary vectors, $\bbw_1\in\{0,1\}^{\bar{E}}$ and $\bbw_2\in\{0,1\}^{\bar{T}}$, with it. An entry of $\bbw_1$ is 1 iff the corresponding column of the full incidence matrix $\bar{\bbB}_1$ identifies an edge present in $\ccalE$, and an entry of $\bbw_2$ is 1 iff the corresponding column of the full incidence matrix $\bar{\bbB}_2$ identifies a triangle in $\ccalT$. 

By reformulating the optimization in terms of these selection vectors $\{\bbw_i\}_{i=1}^2$, enforcing sparsity on $\{\bbw_i\}_{i=1}^2$ naturally yields a sparse SC. Specifically, we can write the quadratic terms $\bbB_i \bbB_i^\top$ as
\begin{equation*}
\bar{\bbB}_i \diag^2(\bbw_i) \bar{\bbB}_i^\top \;=\; \bar{\bbB}_i \diag(\bbw_i)\,\bar{\bbB}_i^\top,
\end{equation*}
which is linear in $\bbw_i$. Note that this recasting also requires the full edge signal matrix $\bar{\bbX}_1 \in \mathbb{R}^{\bar{E}\times P_1}$. Similarly, we will introduce $\bar{\bbX}_0 \in \mathbb{R}^{N\times P_0}$ which is actually the same as ${\bbX}_0$.

To describe idea K2), recall that the SCC states that a triangle $(i,j,k)\in\ccalT$ can only exist if all three of its edges are included in $\ccalE$. To formalize this requirement, we introduce the following lemma.

\begin{mylemma}[Validity condition for an SC]
\label{lem:valid_SC_condition}
Let $\bar{\bbB}_2^+ \in \{0,1\}^{\bar{E} \times \bar{T}}$ denote the (binary) non-oriented edge-to-triangle incidence matrix associated with the complete SC, where $[\bar{\bbB}_2^+]_{l,t} = 1$ indicates that the edge indexed by $l$ belongs to the triangle indexed by $t$. Then, a pair of binary variables $(\bbw_1, \bbw_2)$ defines a valid SC if and only if
\begin{equation}
(\bbone - \bbw_1)^\top \bar{\bbB}_2^+ \bbw_2 = 0.
\label{eq:valid_SC_condition}
\end{equation}
\end{mylemma}

From now on, we will refer to \eqref{eq:valid_SC_condition} as the SCC constraint, which ensures that a triangle $(i,j,k) \in \ccalT$ can only exist if all three of its edges are present in $\ccalE$. The term $\bbw_1$ represents the edge-selection vector and $\bbw_2$ the triangle-selection vector. The product $\bar{\bbB}_2^+ \bbw_2$ yields a nonnegative integer vector, where the non-zero entries indicate which edges are required by the currently selected triangles. Multiplying this vector by $(\bbone - \bbw_1)^\top$ checks for any inconsistencies (i.e., edges that are required by some triangle but not actually selected). 
Hence, the equality in~\eqref{eq:valid_SC_condition} holds if and only if every edge required by any selected triangle is indeed active, guaranteeing that the resulting structure $(\ccalN, \ccalE, \ccalT)$ forms a valid SC. If the equality is violated, at least one triangle is present without all its supporting edges, thus breaking the simplicial closure property.

Leveraging these conventions, we can now reformulate \eqref{E:SC_OPT_v0} as 
\begin{subequations}\label{E:SC_OPT_v1}
\begin{alignat}{2}
	\!\!&\!\min_{ \{\bbw_i\}_{i=1}^2, \{\bar{\bbX}_i\}_{i=0}^1} &&  \alpha_1\|\bbw_1\|_0 \;+\;\alpha_2\|\bbw_2\|_0+
    \varepsilon\|\bar{\bbX}_1\|_F^2
     \nonumber \\
     \!\!&\hspace{9mm} &&+\! \eta_0\| \bar{\bbX}_0 - \bbX_0^\ccalO\|_F^2 + \eta_1\| \bar{\bbTheta} \bar{\bbX}_1 - \bbX_1^\ccalO\|_F^2 \nonumber \\ 
     \!\!&\hspace{9mm} &&+\beta_1\,\tr\bigl(\bar{\bbX}_0\,\bar{\bbX}_0^\top\,\bar{\bbB}_1\,\diag(\bbw_1)\,\bar{\bbB}_1^\top\bigr) \nonumber \\  
	\!\!&\hspace{9mm} &&+  \beta_2\,\tr\bigl(\bar{\bbX}_1 \bar{\bbX}_1^\top\,\bar{\bbB}_2\,\diag(\bbw_2)\,\bar{\bbB}_2^\top\bigr),
 \label{E:SC_OPT_v1_obj} \\ 
	\!\!&\!\hspace{4.5mm} \mathrm{s. \;t. } 
        && \bbw_1 \in \{0,1\}^{\bar{E}},  \;\bbw_2 \in \{0,1\}^{\bar{T}}, 
        \label{E:SC_OPT_v1_c1}\\
    \!\!&\! &&   \|\bbw_1\|_0 \geq E^{\min}, \quad \|\bbw_2\|_0 \geq \blue{T^{\text{budget}}}, 
        \label{E:SC_OPT_v1_c5}\\
	\!\!&\! && 	(\bbone -\bbw_1)^\top \,\bar{\bbB}_2^+\, \bbw_2 = 0, 
        \label{E:SC_OPT_v1_c2}\\
    \!\!&\! && [\bbw_1]_{l}=1 \quad\text{for all } l \in \ccalEobs, 
        \label{E:SC_OPT_v1_c3}
\end{alignat} 
\end{subequations}
where $\bar{\bbTheta}\in\{0,1\}^{E^{\ccalO}\times \bar{E}}$ is the edge sampling matrix selecting the observed edges from all possible edges and $\varepsilon>0$ is an arbitrarily small constant. Expressions~\eqref{E:SC_OPT_v1_obj}--\eqref{E:SC_OPT_v1_c3} mirror \eqref{E:SC_OPT_v0_obj}--\eqref{E:SC_OPT_v0_c3}, while \eqref{E:SC_OPT_v1_c5} (with $E^{\min}>|\ccalEobs|$) prevents the trivial all-zero solution.  \blue{Moreover, the parameter $T^{\text{budget}}$ is a user-specified triangle budget that determines the number of triangles selected during the optimization (exactly $T^{\text{budget}}$ triangles are declared filled; cf. Lemma~\ref{Lemma:GreedyTriangleSelection}) and, consequently, the complexity of the recovered SC.} Finally, the term $\varepsilon\|\bar{\bbX}_1\|_F^2$ in the objective ensures that: i) the optimal solution to the edge interpolation is unique, and ii) such solution sets to zero the signals associated with edges that are neither observed nor involved in active triangles.

Compared to \eqref{E:SC_OPT_v0}, the updated formulation \eqref{E:SC_OPT_v1} is more tractable for four main reasons. 
First, it involves fewer optimization variables, as we now optimize over $(\bbw_1, \bbw_2)$ rather than the larger set of auxiliary variables in \eqref{E:SC_OPT_v0}. 
Second, it contains fewer constraints; for instance, constraint~\eqref{E:SC_OPT_v0_c2} is matrix-valued, whereas the corresponding constraint~\eqref{E:SC_OPT_v1_c2} is scalar. 
Third, the feasible sets of the optimization variables are simpler to describe and handle computationally, since they are defined by binary or box constraints with clear geometric interpretation. 
Fourth, several multilinear terms in \eqref{E:SC_OPT_v0} now reduce to linear expressions in \eqref{E:SC_OPT_v1}, significantly simplifying both the objective and the constraint structure. 

Nonetheless, solving \eqref{E:SC_OPT_v1} remains challenging due to the binary nature of $\{\bbw_i\}_{i=1}^2$ and the bilinear interactions, one arising from the SCC constraint and another from the smoothness term. Next, we discuss an efficient algorithm to address these issues.

\section{A BCD SC approach: Optimal solutions and algorithmic design}\label{S:GreedyAlgorithmAlgorithm}

Our strategy for solving the optimization problem in \eqref{E:SC_OPT_v1} is based on three main steps:
\begin{enumerate}[label=\roman*)]
\item Replace the constraint $(\bbone - \bbw_1)^\top \bar{\bbB}_2^+\, \bbw_2 = 0$ [cf.~\eqref{E:SC_OPT_v1_c2}] with the penalty $\gamma(\bbone - \bbw_1)^\top \bar{\bbB}_2^+\, \bbw_2$ in the objective.
\item Use an alternating optimization scheme to handle the bilinear terms.
\item Apply an (optimal) low-complexity greedy procedure to impose sparsity and address the binary constraints.
\end{enumerate}

Regarding step~i), the components of $(\bbone - \bbw_1)^\top \bar{\bbB}_2^+ \bbw_2$ are all nonnegative, so incorporating it into the objective serves as a valid penalty without requiring any norm. Moreover, this term defines a monotonically nondecreasing function w.r.t. the entries of $\bbw_2$, which is crucial for its role as a penalty: it ensures that any increase in violations of the SCC leads to a larger penalty. In particular, this property guarantees that a filled triangle (2-simplex) is only favored when all three constituent links are present, naturally enforcing consistency with the SC structure.
For step~ii), alternating methods are well suited to problems with bilinear or more generally biconvex structure~\cite{gorski2007biconvex}. Concretely, our approach partitions the variables into four blocks: $\bar{\bbX}_0\in\mathbb{R}^{N\times P_0}$, $\bar{\bbX}_1\in \mathbb{R}^{\bar{E}\times P_1}$, $\bbw_1\in\{0,1\}^{\bar{E}}$, and $\bbw_2\in\{0,1\}^{\bar{T}}$. We then adopt an alternating iterative algorithm, where we optimize over one block of variables at a time, holding the other three blocks fixed.

We analyze each of these four subproblems separately and characterize their optimal solutions. To facilitate the exposition, we first address the subproblem for $\bbw_2$, then turn to $\bbw_1$, and finally discuss the optimization over $\bar{\bbX}_0$ and $\bar{\bbX}_1$.

\subsection{ Optimization of the triangle selection vector}\label{Ss:GreedyAlgorithmAlgorithm_triangles}

We consider $\bar{\bbX}_0$, $\bar{\bbX}_1$ and $\bbw_1$ given and optimize $\bbw_2$. The formal problem is
\begin{alignat}{2}
	\!\!&\!\min_{\bbw_2} \alpha_2\|\bbw_2\|_0+ \beta_2\tr(\bar{\bbX}_1 \bar{\bbX}_1^\top\bar{\bbB}_2\diag(\bbw_2)\bar{\bbB}_2^\top)  
	\label{E:SC_OPT_v1_subp2_c1e} \\ 
	\!\!&\! + \gamma (\bbone\!-\!\bbw_1)^\top \bar{\bbB}_2^+ \bbw_2~\mathrm{\;\;s. \;t. } \, \bbw_2 \!\in\! \{0,1\}^{\bar{T}}\!\!,  \|\bbw_2\|_0 \!\geq \blue{T^{\text{budget}}}\! . \nonumber
\end{alignat} 

Let us write the second  and third terms in the objective as
\begin{align}
	 &\;\tr(\bar{\bbX}_1 \bar{\bbX}_1^\top\bar{\bbB}_2\diag(\bbw_2)\bar{\bbB}_2^\top) \! = \! \textstyle \sum_{t=1}^{\bar{T}}[\bar{\bbB}_2^\top \bar{\bbX}_1 \bar{\bbX}_1^\top\bar{\bbB}_2]_{tt}[\bbw_2]_t, \notag \\
	& (\bbone-\bbw_1)^\top \bar{\bbB}_2^+ \bbw_2= \textstyle \sum_{t=1}^{\bar{T}}[\bar{\bbB}_2^{+\top}(\bbone-\bbw_1)]_{t}[\bbw_2]_t. \notag
\end{align}
These expressions reveal that both cost terms are linear and separable across the entries of $\bbw_2$. This is used to show that the optimal solution to \eqref{E:SC_OPT_v1_subp2_c1e} is given by the next lemma.
\begin{mylemma}\label{Lemma:GreedyTriangleSelection}
	Let us define the triangle-score vector $\bbs_2\in\reals^{\bar{T}}$ as $[\bbs_2]_t=\alpha_2+\beta_2[\bar{\bbB}_2^\top \bar{\bbX}_1 \bar{\bbX}_1^\top\bar{\bbB}_2]_{tt} + \gamma [\bar{\bbB}_2^{+\top}(\bbone-\bbw_1)]_{t}$ and let $\pi_2$ be 
	the permutation function $\{1,...,\bar{T}\} \rightarrow \{1,...,\bar{T}\}$ that orders the elements of $\bbs_2$ in an ascending manner so that $[\bbs_2]_{\pi_2(t)}\leq [\bbs_2]_{\pi_2(t+1)}$. 
 Then, the optimal solution to  \eqref{E:SC_OPT_v1_subp2_c1e} is 
	\begin{eqnarray}\label{E:optimal_greedy_wTriang}
		[\bbw_2]_t
		=\left\{\begin{matrix}	
			1,\;\;&\text{if} \;t\in \{\pi_2(i)\}_{i=1}^{T^{\text{budget}}}\\
			0,\;\;&\text{otherwise}.
		\end{matrix}\right.	
	\end{eqnarray} 
\end{mylemma}
\begin{myproof}
See App. \ref{App:ProofLemmaGreedyTriangleSelection}.
\end{myproof}

The lemma shows that the optimal solution can be obtained via a greedy selection strategy, where each triangle is assigned a score reflecting three contributions: (a) a constant activation cost related to the sparsity of the SC, (b) a smoothness term that measures the influence of triangle activation on the edge signals, and (c) a penalty associated with violations of the simplicial closure condition (SCC). The greedy solution then selects the $\blue{T^{\text{budget}}}$ triangles with the smallest entries in the score vector $\bbs_2$, declaring them as filled. Importantly, the SCC penalty  vanishes when all three edges of a triangle are already present in $\bbw_1$. This ensures that a 2-simplex is only favored when its constituent links exist, thereby enforcing the necessary structural consistency with the SC model. While omitting the constant term does not affect the ranking of triangles, keeping it becomes relevant when replacing the $\ell_0$ penalty with (reweighted) $\ell_1$ relaxations. Finally, since the scores in $\bbs_2$ are always nonnegative, ties across triangles can be broken arbitrarily without affecting optimality.

From a computational perspective, sorting operations have an average complexity of $O(\bar{T}\log \bar{T})$, and thus are relatively efficient. However, computing the scores requires additional complexity, with the most expensive operation associated with the computation of the smoothness term.

\subsection{Optimization of the edge selection vector}\label{Ss:GreedyAlgorithmAlgorithm_edges}

Here, we consider $\bar{\bbX}_0$, $\bar{\bbX}_1$ and $\bbw_2$ given and optimize $\bbw_1$. The resultant problem is  
\begin{alignat}{2}
	\!\!&\!\min_{\bbw_1} &&   \alpha_1\|\bbw_1\|_0+ \beta_1\tr(\bar{\bbX}_0\bar{\bbX}_0^\top\bar{\bbB}_1\diag(\bbw_1)\bar{\bbB}_1^\top) \nonumber \\ 
    \!\!&\! && + \gamma (\bbone-\bbw_1)^\top \bar{\bbB}_2^+ \bbw_2  \label{E:SC_OPT_v1_subp1_obj} \\ 
	\!\!&\! \mathrm{\;\;s. \;t.\, } && \bbw_1 \in \{0,1\}^{\bar{E}},  \; [\bbw_1]_{l}=1 \;\; \forall \;l\in \ccalEobs, \|\bbw_1\|_0 \geq E^{\min} . \nonumber
\end{alignat} 

As before, we write the second and third terms as
\begin{align}
	&\tr(\bar{\bbX}_0\bar{\bbX}_0^\top\bar{\bbB}_1\diag(\bbw_1)\bar{\bbB}_1^\top) = \sum_{l=1}^{\bar{E}}[\bar{\bbB}_1^\top\bar{\bbX}_0\bar{\bbX}_0^\top\bar{\bbB}_1]_{ll}[\bbw_1]_l,  \notag \\
	&(\bbone-\bbw_1)^\top \bar{\bbB}_2^+ \bbw_2 = \sum_{l=1}^{\bar{E}}[\bar{\bbB}_2^+\bbw_2]_{l}-\sum_{l=1}^{\bar{E}}[\bar{\bbB}_2^+\bbw_2]_{l}[\bbw_1]_l. \notag
\end{align}

A conceptual point arises when considering the role of the SCC structure. In the previous subsection, triangles are identified via $\bbw_2$, and in principle, all edges forming a triangle should be included. One may ask: should we explicitly enforce certain entries of $\bbw_1$ to $1$ whenever a triangle is found in $\bbw_2$, and then optimize the remaining pairwise connections in \eqref{E:SC_OPT_v1_subp1_obj}?  
From a mathematical perspective, explicitly enforcing SCC as a hard constraint can be very restrictive. If such constraints are active at every iteration, only pre-existing triangles can be created/filled, and once a triangle is declared/filled, the corresponding links cannot be removed. This significantly limits the flexibility of block-coordinate updates and may invalidate convergence guarantees provided by standard BCD theory.  

On the other hand, augmenting the objective with the SCC penalty term (via $\gamma$) provides a practical compromise:  
\begin{itemize}
    \item For sufficiently large $\gamma$, the SCC constraint is approximately enforced.  
    \item The BCD algorithm retains degrees of freedom to explore alternative solutions, which allows convergence analysis to remain valid.  
    \item \blue{The main trade-off is that the continuous solution may not be strictly feasible w.r.t the SCC constraints, since a finite penalty does not theoretically guarantee exact feasibility. In practice, however, for sufficiently large values of $\gamma$, the recovered solution satisfies the SCC. To guarantee that the final output always corresponds to a valid simplicial complex, a final post-processing step is applied: if any filled triangle is present without all of its incident edges, the missing edges are added, thereby enforcing the simplicial closure constraint.} 
\end{itemize}
Thus, one can prioritize either strict SCC enforcement or algorithmic flexibility. Our approach follows the latter, treating SCC as a soft constraint to maintain theoretical guarantees and practical convergence, \blue{relying on a final rounding step to ensure strict SCC satisfaction.}

The main difference relative to problem \eqref{E:SC_OPT_v1_subp2_c1e} is that here some of the links are known. As a result, the counterpart to Lemma~\ref{Lemma:GreedyTriangleSelection} for the edge-selection vector is given next.
\begin{mylemma}\label{Lemma:GreedyEdgeSelection}
	Let us define the edge-score vector $\bbs_1\in\reals^{\bar{E}}$ as 
 $[\bbs_1]_l=-1$ if $l \in \ccalEobs$ and
 $[\bbs_1]_l=\alpha_1+\beta_1[\bar{\bbB}_1^\top\bar{\bbX}_0\bar{\bbX}_0^\top\bar{\bbB}_1]_{ll} - \gamma [\bar{\bbB}_2^+\bbw_2]_{l}$ otherwise and let $\pi_1$ be 
the permutation function $\{1,...,\bar{E}\} \rightarrow \{1,...,\bar{E}\}$ that orders the elements of $\bbs_1$ in an ascending manner so that $[\bbs_1]_{\pi_1(l)}\leq [\bbs_1]_{\pi_1(l+1)}$. Additionally, let $E^{\mathrm{neg}}$ denote the number of entries of $\bbs_1$ that are negative and define $E^{\mathrm{act}}=\max\{ E^{\mathrm{neg}},E^{\min} \}$. Then, the optimal solution to  \eqref{E:SC_OPT_v1_subp1_obj} is 
	\begin{eqnarray}
	[\bbw_1]_l
	=\left\{\begin{matrix}	
		1,\;\;&\text{if} \;l\in \{\pi_1(i)\}_{i=1}^{E^{\mathrm{act}}}\\
		0,\;\;&\text{otherwise}.
		\end{matrix}\right.	
	\end{eqnarray} 
\end{mylemma}
\begin{myproof}
See App. \ref{App:ProofLemmaGreedyEdgeSelection}
\end{myproof}

Once again, these findings indicate that greedily activating edges (according to the score defined in the lemma) is optimal. First, since all observed edges yield a negative score, we are guaranteed that $[\bbw_1]_l = 1$ for all $l \in \ccalEobs$. Second, in contrast to the solution for the optimal triangle selection vector, the number of activated edges can exceed $E^{\min}$. This situation arises, for instance, when the vector $\bbw_2$ contains a large number of triangles and the penalty parameter $\gamma$ is sufficiently large.

\subsection{Optimization of the nodal signals}\label{Ss:GreedyAlgorithmAlgorithm_nodalsignals}

Here we consider $\bbw_1$, $\bbw_2$ and $\bar{\bbX}_1$ given and optimize over $\bar{\bbX}_0$. The resultant problem is  
\vspace{-.1cm}
\begin{alignat}{2}
	\!\!&\!\min_{\bar{\bbX}_{0}} &&  \; \beta_1\tr(\bar{\bbX}_0 \bar{\bbX}_0^\top\bar{\bbB}_1\diag(\bbw_1)\bar{\bbB}_1^\top) + \eta_0\| \bar{\bbX}_0-\bbX_0^\ccalO\|_F^2  \label{E:SC_OPT_v1_subp3_obj} 
\end{alignat} 
The problem is convex and differentiable, and its solution is provided in the next lemma.
\vspace{-0.25cm}
\begin{mylemma}\label{Lemma:NodalSignalsDenoised}
\vspace{-0.105cm} The denoised nodal signals are given by
\begin{equation}\label{E:NodalSignalsDenoised}
	\bar{\bbX}_0=  \Big(\bbI + \frac{\beta_1}{\eta_0}\bar{\bbB}_1\diag(\bbw_1)\bar{\bbB}_1^\top\Big)^{-1} \bbX_0^\ccalO.
\end{equation}
\end{mylemma}
\begin{myproof}
When the topology of the SC is given, the denoising of the nodal signals is just a least-squares problem, which can be solved by setting the gradient to zero. Specifically, note that the gradient of the objective in \eqref{E:SC_OPT_v1_subp3_obj}  w.r.t. $\bar{\bbX}_{0}$ is 
\begin{eqnarray}\label{E:gradient_wrt_X0}
    2\beta_1 \bar{\bbB}_1\diag(\bbw_1)\bar{\bbB}_1^\top\bar{\bbX}_0 + 2\eta_0(\bar{\bbX}_0-\bbX_0^\ccalO) 
\end{eqnarray}
where, for convenience, the gradient has been written in a matrix form. Setting \eqref{E:gradient_wrt_X0} to zero and solving w.r.t. $\bar{\bbX}_{0}$ yields
\begin{eqnarray}\label{E:gradient_wrt_X0_set_to_zero}
    \big(\beta_1\bar{\bbB}_1\diag(\bbw_1)\bar{\bbB}_1^\top + \eta_0\bbI\big)\bar{\bbX}_0=\eta_0\bbX_0^\ccalO. 
\end{eqnarray}
The matrix on the LHS is a positive definite matrix. Hence, its inverse exists and, as a result, the optimal expression given in the lemma follows from \eqref{E:gradient_wrt_X0_set_to_zero}.
\end{myproof}

The structure of the solution in Lemma~\ref{Lemma:NodalSignalsDenoised} mirrors that of many (regularized) denoising problems, with the ratio $\beta_1/\eta_0$ governing the trade-off between smoothness and fidelity to the observed data.

\subsection{Optimization of the edge signals}\label{Ss:GreedyAlgorithmAlgorithm_edgesignals}

In this step, we consider $\bbw_1$, $\bbw_2$, and $\bar{\bbX}_0$ given, and optimize over the edge signals $\bar{\bbX}_1$. The resulting problem is  
\vspace{-.1cm}
\begin{alignat}{2}
	\!\!&\!\min_{\bar{\bbX}_{1}} &&  \; \beta_2\tr(\bar{\bbX}_1 \bar{\bbX}_1^\top\bar{\bbB}_2\diag(\bbw_2)\bar{\bbB}_2^\top) \!+\! \eta_1\|\bar{\bbTheta} \bar{\bbX}_1\!-\!\bbX_1^\ccalO\|_F^2
   \notag  \\ & && +
    \varepsilon\|\bar{\bbX}_1\|_F^2,  \label{E:SC_OPT_v1_subp4_obj} 
\end{alignat} 
which is convex and smooth and whose solution is given next.

\vspace{-0.25cm}
\begin{mylemma}\label{Lemma:EdgeSignalsInterp}
The interpolated edge signals are given by
\begin{equation}\label{E:EdgeSignalsInterp}
	\bar{\bbX}_1=  \Big(\blue{\frac{\varepsilon}{\eta_1}\bbI}+\bar{\bbTheta}^\top \bar{\bbTheta} + \frac{\beta_2}{\eta_1}\bar{\bbB}_2\diag(\bbw_2)\bar{\bbB}_2^\top \Big)^{-1} \bar{\bbTheta}^\top\bbX_1^\ccalO.
\end{equation}
\end{mylemma}
\begin{myproof}
The proof follows directly from the first-order optimality conditions and mirrors the derivation in Lemma~\ref{Lemma:NodalSignalsDenoised}.
\end{myproof}

Several remarks are in order. First, multiplying the observed edge signals $\bbX_1^\ccalO$ by $\bar{\bbTheta}^\top$ effectively performs zero-padding over unobserved edges. Second, since $\varepsilon$ is an arbitrarily small constant, the primary role of the inverse in \eqref{E:EdgeSignalsInterp} is to balance data fidelity (through $\bar{\bbTheta}^\top \bar{\bbTheta}$) with the requirement that edge signals remain smooth across filled triangles (via $\bar{\bbB}_2 \diag(\bbw_2)\bar{\bbB}_2^\top$). In this way, the solution directly links edge-signal estimation to the inferred simplicial topology. This trade-off is governed by the ratio $\tfrac{\beta_2}{\eta_1}$: small values prioritize agreement with the observations, whereas larger values enforce stronger smoothness in the curl space, which is particularly important in noisy settings. Third, the role of $\varepsilon$ is critical for edges that are neither observed nor participate in any filled triangle. To see this, note that the matrix $\bar{\bbB}_2 \diag(\bbw_2)\bar{\bbB}_2^\top$ is sparse, with rows and columns vanishing for edges that do not belong to any filled triangle. Similarly, $\bar{\bbTheta}^\top \bar{\bbTheta}$ has zero rows and columns corresponding to unobserved edges. Setting $\varepsilon = 0$ would therefore render the matrix in \eqref{E:EdgeSignalsInterp} rank-deficient, leading to multiple solutions. Consequently, the inclusion of the term $\blue{\frac{\varepsilon}{\eta_1}\bbI}$ is essential, as it a) ensures that the matrix is invertible, and hence the solution is unique, and b) sets to zero the rows of $\bar{\bbX}_1$ corresponding to edges that are neither observed nor involved in any filled triangle.


Overall, the interpolation in \eqref{E:EdgeSignalsInterp} should not be viewed as a generic denoising or inpainting step, but as a mechanism that integrates observed edge signals with topological constraints imposed by the SC. This highlights the key difference with classical edge-signal interpolation: here, the smoothness prior is coupled with the inference of the simplicial topology.

\subsection{BCD SC-learning} 
After detailing the solution of each subproblem, this subsection presents the iterative procedure used to learn the SC topology from smooth observed SC signals. We summarize the proposed greedy algorithm and analyze its computational complexity, highlighting its advantages over generic convex optimization approaches and discussing further reductions enabled by iterative signal updates.

Let $\ell=1,...,\ell^{\max}$ denote the iteration index and let $(\bbw_1^{(\ell)},\bbw_2^{(\ell)},\bar{\bbX}_0^{(\ell)},\bar{\bbX}_1^{(\ell)})$ be the solution estimated at iteration $\ell$. The steps to learn the topology of the SC from the smooth SC signals $\bbX_0^\ccalO$ and $\bbX_1^\ccalO$ are summarized in Alg. \ref{Alg:GreedySCL}, which is labeled as ``GreedySCL''. 
The final estimates are $\hbw_1=\bbw_1^{(\ell^{\max})}$,  $\hbw_2=\bbw_2^{(\ell^{\max})}$, $\hat{\bar{\bbX}}_0=\bar{\bbX}_0^{(\ell^{\max})}$, and $\hat{\bar{\bbX}}_1=\bar{\bbX}_1^{(\ell^{\max})}$.

\RestyleAlgo{ruled}
\begin{algorithm}[t]
	\small
	\SetKwInOut{Output}{Outputs}
	\SetKwInput{KwData}{Input}
	\KwData{$\bbX_0^{\ccalO}$, $\bbX_1^{\ccalO}$, $\ccalE^\ccalO$, $E^{\min}$, $\blue{T^{\text{budget}}}$, $\ell^{\max}$, $\delta$}
	\vspace{0.1cm}
	\Output{$\hbw_1$, $\hbw_2$, $\hat{\bar{\bbX}}_0$ and $\hat{\bar{\bbX}}_1$}
	\SetAlgoLined
	\vspace{0.1cm}
	
	\blue{Initialize $\bar{\bbX}_0^{(0)} = \bbX_0^{\ccalO}$ and $\bar{\bbX}_1^{(0)} = \bar{\bbTheta}^\top \bbX_1^{\ccalO}$} \\
	Initialize $\bbw_1^{(0)} = \bb0$ and $\bbw_2^{(0)} = \bb0$ \\
	\blue{Initialize $\ell^{\mathrm{end}}=0$.} \\
	\vspace{0.1cm}
	
	\For{$\ell=0$ \KwTo $\ell^{\max}-1$ }{
		
		Update \!\!\! $\bbw_1^{(\ell+1)}$ \!\!\!\! using \!\!\! Lemma \! \ref{Lemma:GreedyEdgeSelection} \!\!\! with \!\!\! $\bbw_2=\bbw_2^{(\ell)}$ and $\bar{\bbX}_0^{(\ell)}$.\\
		\vspace{0.1cm}
		
		Update \!\!\! $\bbw_2^{(\ell+1)}$ \!\!\!\! using \!\!\! Lemma \! \ref{Lemma:GreedyTriangleSelection} \!\!\! with \!\!\! $\bbw_1=\bbw_1^{(\ell+1)}$ and $\bar{\bbX}_1=\bar{\bbX}_1^{(\ell)}$.\\
		\vspace{0.1cm}

		Update \!\!\! $\bar{\bbX}_0^{(\ell+1)}$ \!\!\!\! using \!\!\! Lemma \!\ref{Lemma:NodalSignalsDenoised} \!\!\! with \!\!\! $\bbw_1=\bbw_1^{(\ell+1)}$.\\
		\vspace{0.1cm}
        
		Update \!\!\! $\bar{\bbX}_1^{(\ell+1)}$ \!\!\!\! using \!\!\! Lemma \!\ref{Lemma:EdgeSignalsInterp} \!\!\! with \!\!\! $\bbw_2=\bbw_2^{(\ell+1)}$.\\
		\vspace{0.1cm}

		\blue{$\ell^{\mathrm{end}}=\ell+1$.}\\
		\vspace{0.1cm}
		
		\If{$\|\bar{\bbX}_0^{(\ell+1)}-\bar{\bbX}_0^{(\ell)}\|_F^2+\|\bar{\bbX}_1^{(\ell+1)}-\bar{\bbX}_1^{(\ell)}\|_F^2+\|\bbw_1^{(\ell+1)}-\bbw_1^{(\ell)}\|_F^2+\|\bbw_2^{(\ell+1)}-\bbw_2^{(\ell)}\|_F^2 \le \delta$}{
			\textbf{break};
		}
	}
	\vspace{0.1cm}

	\blue{$\hbw_2=\bbw_2^{(\ell^{\mathrm{end}})}$,}
	\blue{$\hbw_1=\bbw_1^{(\ell^{\mathrm{end}})}$,}
	\blue{$\hat{\bar{\bbX}}_0=\bar{\bbX}_0^{(\ell^{\mathrm{end}})}$,
	$\hat{\bar{\bbX}}_1=\bar{\bbX}_1^{(\ell^{\mathrm{end}})}$.}
    
		\blue{Add to $\hbw_1$ all edges required by the filled triangles in $\hbw_2$.}\\
	\caption{SC-learning algorithm from smooth SC signals (GreedySCL).}
	\label{Alg:GreedySCL}
\end{algorithm}

To investigate the computational complexity of Alg.~\ref{Alg:GreedySCL}, note that, given the discrete nature of the optimization and the large number of variables (recall that $\bar{T}$, the length of $\bbw_2$, scales with $N^3$), complexity is kept under control due to the greedy nature of the solution. Sorting the edges and triangles takes $O(\bar{E}\log(\bar{E})) = O(N^2\log(N))$ and $O(\bar{T}\log(\bar{T})) = O(N^3\log(N))$, respectively. Alg.~\ref{Alg:GreedySCL} requires running $\ell^{\max}$ iterations, but this term is modest in practice (typically a few tens), keeping the overall computational cost manageable. \blue{Although the number of candidate triangles grows cubically with the number of nodes, the proposed method avoids exhaustive combinatorial search by relying on greedy ranking and selection operations. Consequently, the dominant computational cost scales with sorting and matrix operations, making the algorithm practical for considerably larger graphs than those that could be handled by a brute-force search.}

If the SC-learning problem were convex, a generic solver would require a computational complexity that scales with the power $3.5$ of the number of variables, leading to complexities dominated by the terms $O(\bar{E}^{3.5}) = O(N^7)$ and $O(\bar{T}^{3.5}) = O(N^{10.5})$. In contrast, the proposed greedy formulation avoids these high-order dependencies, resulting in substantial computational savings. 

Shifting focus to the node and edge signals, complexity can also be reduced. In particular, since Alg.~\ref{Alg:GreedySCL} adopts an iterative strategy, rather than finding the optimal $\bar{\bbX}_0$ and $\bar{\bbX}_1$ for each iteration $\ell$ using the (pseudo-)inverses in \eqref{E:NodalSignalsDenoised} and \eqref{E:EdgeSignalsInterp}, one can approximate the least-squares solution by, e.g., running a few iterations of gradient descent.
Section~\ref{Sec:NumericalResults} will explore this alternative in different numerical setups, showing that the associated optimality loss is negligible. 
\subsection{Convergence results}

This section proves that, under suitable conditions, Alg.~\ref{Alg:GreedySCL} converges to a \blue{coordinate-wise minimum} that is a stationary point. The key steps to show this are: i) relaxing the norm-zero terms and the binary constraints in  \eqref{E:SC_OPT_v1} to linear ones, ii) implementing a BCD approach for the relaxed problem, iii) showing that, under certain conditions, the solution to the relaxed block minimization is similar to that of the binary block optimization (i.e., to the greedy optimization provided by Alg. \ref{Alg:GreedySCL}), and iv) leveraging the convergence results of \cite{tseng2001convergence}, which deals with block-convex problems.  

We begin by formally stating the following assumption:
\textit{\begin{enumerate}[label=(AS5)]
    \item  The iterates generated by Alg.~\ref{Alg:GreedySCL} are such that, at every iteration $\ell$, the score vectors satisfy:
\begin{enumerate}
    \item \blue{$[\bbs_1^{\ell}]_{\pi_1(E^{\mathrm{act}})} \neq [\bbs_1^{\ell}]_{\pi_1(E^{\mathrm{act}}+1)}$ and $[\bbs_1^{\ell}]_{\pi_1(E^{\mathrm{act}}+1)} \neq 0$, with $E^{\mathrm{act}}=\max\{E^{\mathrm{neg}},E^{\min}\}$ computed from $\bbs_1^{\ell}$ as in Lemma~\ref{Lemma:GreedyEdgeSelection}; and} 
    \item $[\bbs_2^{\ell}]_{\pi_2(\blue{T^{\text{budget}}})} \neq [\bbs_2^{\ell}]_{\pi_2(\blue{T^{\text{budget}}}+1)}$.
\end{enumerate}
\end{enumerate}}
\noindent This assumption is reasonable because, in practice, the score vectors generated by greedy updates typically do not produce exact ties \blue{or exactly-zero scores} due to randomness in initialization or small variations in the data. If ties occur, they can be resolved consistently without affecting the convergence behavior of the algorithm. \blue{Two comments on condition (a) are in order. First, the condition is stated at the active cut $E^{\mathrm{act}}$ rather than at $E^{\min}$ because, whenever $E^{\mathrm{neg}} > E^{\min}$, the cardinality constraint is inactive and the set of selected edges is determined by the sign of the scores rather than by the cardinality cut. Second, the requirement $[\bbs_1^{\ell}]_{\pi_1(E^{\mathrm{act}}+1)} \neq 0$ is needed because an unobserved edge whose score is exactly zero contributes nothing to the objective of the (relaxed) edge subproblem, so that the associated entry could take any value in $[0,1]$ without altering the objective, giving rise to multiple optimal solutions. Note that, when $E^{\mathrm{neg}} < E^{\min}$, a zero score can only be problematic if it generates a tie at the cut (a zero score below the cut is harmless, since the corresponding edge is forced to be active by the cardinality constraint), a case already excluded by the first requirement. Finally, no zero-score condition is needed in (b), since all triangle scores are bounded below by $\alpha_2 > 0$.}

Under this assumption, it readily follows that the greedy solutions in Lemmas \ref{Lemma:GreedyTriangleSelection} and \ref{Lemma:GreedyEdgeSelection} are unique. Furthermore, we formulate the following continuous version of \eqref{E:SC_OPT_v1}.  
\begin{subequations}\label{E:SC_OPT_v1_relaxed}
\begin{alignat}{2}
	\!\!&\!\min_{ \{\bbw_i\}_{i=1}^2, \{\bar{\bbX}_i\}_{i=0}^1} &&  \gamma(\bbone -\bbw_1)^\top \bar{\bbB}_2^+ \bbw_2+\alpha_1\|\bbw_1\|_1+\alpha_2\|\bbw_2\|_1
     \nonumber \\
     \!\!&\hspace{9mm} &&+\varepsilon\|\bar{\bbX}_1\|_F^2 \!+\!\eta_0\| \bar{\bbX}_0\!-\!\!\bbX_0^\ccalO\|_F^2+\! \eta_1\| \bar{\bbTheta} \bar{\bbX}_1\!-\!\!\bbX_1^\ccalO\|_F^2   \nonumber \\ 
     \!\!&\hspace{9mm} &&+\blue{\beta_1\tr(\bar{\bbX}_0\bar{\bbX}_0^\top\bar{\bbB}_1\diag(\bbw_1)\bar{\bbB}_1^\top)} \nonumber \\  
	\!\!&\hspace{9mm} &&+  \beta_2\tr(\bar{\bbX}_1 \bar{\bbX}_1^\top\!\bar{\bbB}_2\diag(\bbw_2)\bar{\bbB}_2^\top) \!  \label{E:SC_OPT_v1_obj_relaxed} \\ 
	\!\!&\!\hspace{4.5mm} \mathrm{\;\;s. \;t. } && \bbw_1 \in [0,1]^{\bar{E}},  \;\bbw_2 \in [0,1]^{\bar{T}}, \label{E:SC_OPT_v1_c1_relaxed}\\
	\!\!&\! && [\bbw_1]_{l}=1 \;\;\blue{\text{for all}} \;l\in \ccalEobs,\;\; \label{E:SC_OPT_v1_c3_relax3ed}\\  
	\!\!&\! &&   \!\sum_{l=1}^{\bar{E}} [\bbw_1]_l \!\geq\! E^{\min} \;  \text{ and } \; \!\sum_{l=1}^{\bar{T}} [\bbw_2]_l \!\geq\! \blue{T^{\text{budget}}}. \label{E:SC_OPT_v1_c5_relaxed}
\end{alignat} 
\end{subequations}
Compared to \eqref{E:SC_OPT_v1}, the primary modifications in \eqref{E:SC_OPT_v1_relaxed} are:
a) as foreshadowed after \eqref{E:SC_OPT_v1}, the SCC constraint in \eqref{E:SC_OPT_v1_c2} is replaced by the penalty term $\gamma(\bbone - \bbw_1)^\top \bar{\bbB}_2^+ \bbw_2$; b) the zero-norms in the objective are replaced by one-norms; c) the binary constraints $\bbw_1 \in \{0,1\}^{\bar{E}}$ and $\bbw_2 \in \{0,1\}^{\bar{T}}$ are relaxed to $\bbw_1 \in [0,1]^{\bar{E}}$ and $\bbw_2 \in [0,1]^{\bar{T}}$; and d) the constraints $\|\bbw_1\|_0 \geq E^{\min}$ and $\|\bbw_2\|_0 \geq \blue{T^{\text{budget}}}$ in \eqref{E:SC_OPT_v1_c5} are replaced by the linear inequalities $\sum_{l=1}^{\bar{E}} [\bbw_1]_l \geq E^{\min}$ and $\sum_{l=1}^{\bar{T}} [\bbw_2]_l \geq \blue{T^{\text{budget}}}$ in \eqref{E:SC_OPT_v1_c5_relaxed}. Because $[\bbw_1]_l$ and $[\bbw_2]_t$ lie in $[0,1]$, satisfying $\sum_{l=1}^{\bar{E}} [\bbw_1]_l \geq E^{\min}$ guarantees that $\|\bbw_1\|_0 \geq E^{\min}$ (an analogous statement holds for $\bbw_2$).

Building on these preliminaries, we can now establish the convergence of Alg.~\ref{Alg:GreedySCL}.
\begin{theorem}\label{Theo:ConvergenceBlockCoordinate}
    Let $f(\bbw_1,\bbw_2,\bar{\bbX}_0,\bar{\bbX}_1)$ be the objective function in \eqref{E:SC_OPT_v1_obj_relaxed}, and let $\ccalZ^*$ denote the set of stationary points of \eqref{E:SC_OPT_v1_relaxed}. Denote by
    \[
    \bbz^{(\ell)} \;=\; \bigl[\bbw_1^{(\ell)},\;\bbw_2^{(\ell)},\;\vvec(\bar{\bbX}_0^{(\ell)})^\top,\;\vvec(\bar{\bbX}_1^{(\ell)})^\top\bigr]^\top
    \]
     the solution produced by Alg.~\ref{Alg:GreedySCL} after $\ell$ iterations. Under assumption (AS5), it holds that $\bbz^{(\ell)}$ converges to \blue{the set $\ccalZ^*$ of stationary points of the relaxed penalized problem} \eqref{E:SC_OPT_v1_relaxed} as $\ell \to \infty$, i.e.,
    \begin{equation}
         \lim_{\ell\to\infty} \mathsf{d}(\bbz^{(\ell)}~|\ccalZ^*) = 0, 
        \label{E:convergenceTheorem}
    \end{equation}
    where $\mathsf{d}\bigl(\bbz \mid \ccalZ^*\bigr) := \min_{\bbz^* \in \ccalZ^*} \|\bbz - \bbz^*\|_2$.
\end{theorem}

\begin{myproof}
See App.~\ref{App:ProofTheoConvergenceBlockCoordinate}.
\end{myproof}

The proof unfolds in two steps. First, we employ a block alternating minimization scheme for \eqref{E:SC_OPT_v1_relaxed} and show that the solutions in Lemmas~\ref{Lemma:GreedyTriangleSelection}--\ref{Lemma:EdgeSignalsInterp}, originally tailored to \eqref{E:SC_OPT_v1}, also solve the subproblems associated with \eqref{E:SC_OPT_v1_relaxed}. 
Second, we demonstrate that \eqref{E:SC_OPT_v1_relaxed} satisfies the conditions of \cite[Th.~4]{tseng2001convergence}, which implies \eqref{E:convergenceTheorem}. 
As detailed in the proof, the results in \cite{tseng2001convergence} ensure convergence to a block-coordinate minimum, and due to the regularity of the objective in \eqref{E:SC_OPT_v1_obj_relaxed}, this point is also stationary. 

The convergence guarantee applies to the relaxed problem in  \eqref{E:SC_OPT_v1_relaxed}, while the original discrete formulation in \eqref{E:SC_OPT_v1} does not admit a well-defined notion of stationarity. \blue{Nonetheless, the iterates generated by Alg.~\ref{Alg:GreedySCL} are binary, so they are feasible for \eqref{E:SC_OPT_v1_relaxed} as well as for the penalized counterpart of \eqref{E:SC_OPT_v1}, i.e., the version of \eqref{E:SC_OPT_v1} where the SCC constraint \eqref{E:SC_OPT_v1_c2} is replaced with the penalty $\gamma(\bbone - \bbw_1)^\top \bar{\bbB}_2^+ \bbw_2$. Moreover, since we operate in a multiconvex setting with BCD, each iteration does not increase the objective w.r.t. the updated block. In contrast, feasibility w.r.t. the SCC constraint \eqref{E:SC_OPT_v1_c2} is not guaranteed along the iterations; as discussed next, simplicial closure is promoted by the penalty term and, if needed, enforced by a final closure step.}

\blue{From a practical perspective, the convergence guarantee applies to the relaxed penalized formulation, where the SCC is promoted through the penalty parameter $\gamma$. Although the convergence result does not by itself guarantee exact satisfaction of the SCC, in practice we observe that sufficiently large values of $\gamma$ drive the iterates to solutions that already satisfy the simplicial closure constraint. To guarantee that the final output always corresponds to a valid simplicial complex, we perform a final closure step after convergence: if any selected filled triangle is missing one or more of its supporting edges, the corresponding edges are activated. Therefore, the estimated simplicial complex always satisfies the SCC, while preserving the convergence properties of the underlying relaxed optimization.}

\begin{remark}\label{R:closure_bound}
\blue{The objective increase incurred by the final closure step can be
bounded explicitly. At convergence, flipping a non-selected edge entry
$[\bbw_1]_l$ from $0$ to $1$ changes the objective of the relaxed penalized
problem in \eqref{E:SC_OPT_v1_relaxed} by exactly
$[\bbs_1]_l=\alpha_1+\beta_1[\bar{\bbB}_1^\top\bar{\bbX}_0\bar{\bbX}_0^\top\bar{\bbB}_1]_{ll}
-\gamma[\bar{\bbB}_2^{+}\bbw_2]_{l}$, which is nonnegative for every
non-selected edge (an edge with a negative score would have been activated by
the greedy update in Lemma~\ref{Lemma:GreedyEdgeSelection}). Hence, denoting
by $\ccalE^{\mathrm{cl}}$ the set of edges added by the closure step, the
total objective increase equals $\sum_{l\in\ccalE^{\mathrm{cl}}}[\bbs_1]_l$,
with $|\ccalE^{\mathrm{cl}}|$ being at most three times the number of
SCC-violating triangles. This quantifies the (typically small) price paid for
guaranteeing that the final output is a valid SC.}
\end{remark}

\section{Numerical results}\label{Sec:NumericalResults}

This section is organized into four parts. First, we evaluate the performance of the proposed approach for topology recovery under different experimental settings. Second, we assess its robustness in denoising and recovering both node and edge signals. Third, we analyze algorithmic aspects, including convergence behavior and computational complexity. Finally, we demonstrate the effectiveness and practical utility on a real co-authorship dataset, comparing its performance against relevant baselines.\footnote{Due to space limitations, we present here only a selection of the tests run. See the GitHub repository (\url{https://github.com/andreibuciulea/SC_Learning}) for full details on the simulation setup and additional experiments.}

Unless stated otherwise, the synthetic experiments follow a common setup. We consider three types of random graphs: Erdős–Rényi (ER) with connection probability $0.3$, stochastic block model (SBM) with 4 communities and intra- and inter-community connection probabilities of $0.8$ and $0.2$, respectively, and Barabási–Albert (BA) graphs where each new node attaches to $m=3$ existing nodes. All graphs contain $N=20$ nodes, $50\%$ of all the triangles are filled, and $70\%$ of the edge signals are observed. Node and edge signals, denoted $\bbX_0$ and $\bbX_1$, are synthetically generated with $P_0 = P_1 = 10^3$ samples. 
Results are averaged over 100 independent SC realizations. The compared methods include: ``G-SCL'', our proposed approach summarized in Alg. \ref{Alg:GreedySCL} ; ``S-SCL'', which uses a greedy algorithm to estimate $\bbB_1$ and $\bbB_2$ from $\bbX_0$ and $\bbX_1$ without enforcing any relationship between them; and ``RC''\cite{zomorodian2010fast}, which estimates $\bbB_1$ and $\bbB_2$ based on the correlation structure of the node signals.

\blue{The hyperparameters of all methods are selected through a grid search, choosing the combination that yields the best average performance on the corresponding experimental setting. Unless otherwise specified, the same tuning strategy is adopted throughout all experiments and applied to every method, so that the comparison is fair. For reproducibility, the complete implementation, including the hyperparameter configurations used in all experiments, is publicly available in the code repository.}

\subsection{Topology recovery}




\begin{figure*}[h]
    \centering

    \begin{subfigure}[b]{0.302\textwidth}
        \includegraphics[width=\textwidth]{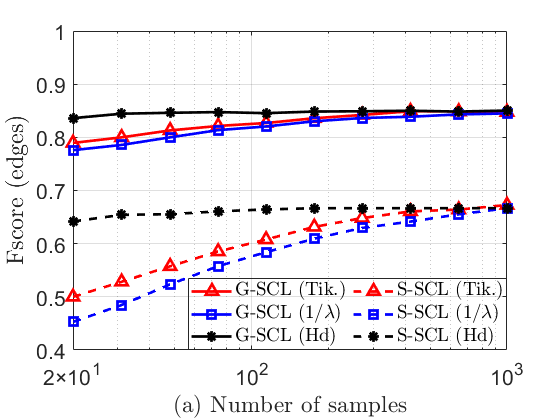}
    \end{subfigure}
    \begin{subfigure}[b]{0.302\textwidth}
        \includegraphics[width=\textwidth]{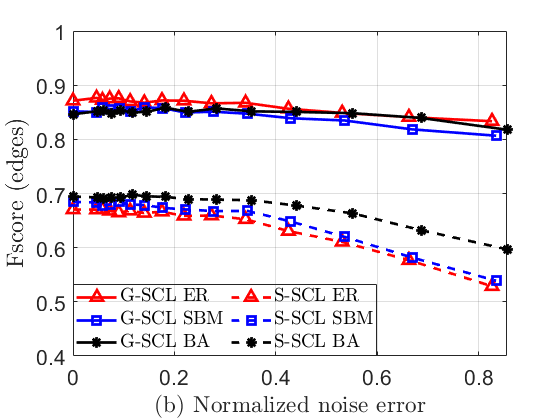}
    \end{subfigure}
    \begin{subfigure}[b]{0.302\textwidth}
        \includegraphics[width=\textwidth]{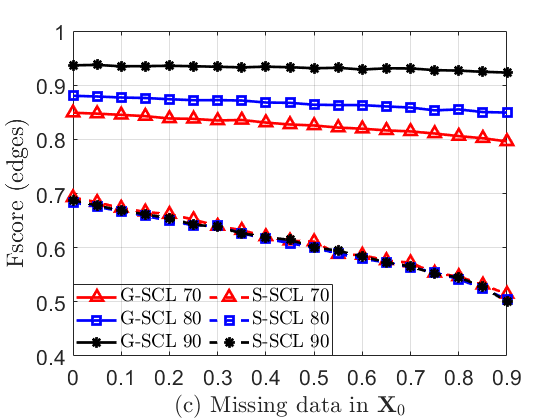}
    \end{subfigure}

    \vspace{0.103cm} 

    \begin{subfigure}[b]{0.302\textwidth}
        \includegraphics[width=\textwidth]{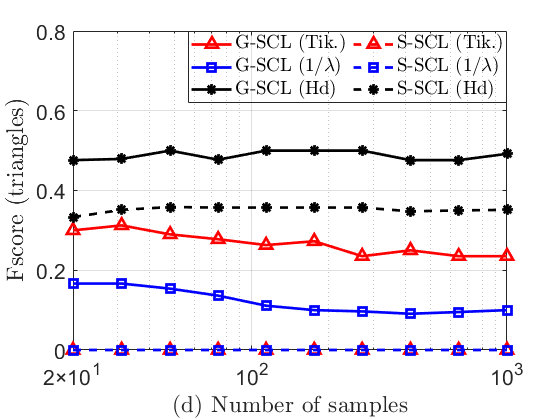}
    \end{subfigure}
    \begin{subfigure}[b]{0.302\textwidth}
        \includegraphics[width=\textwidth]{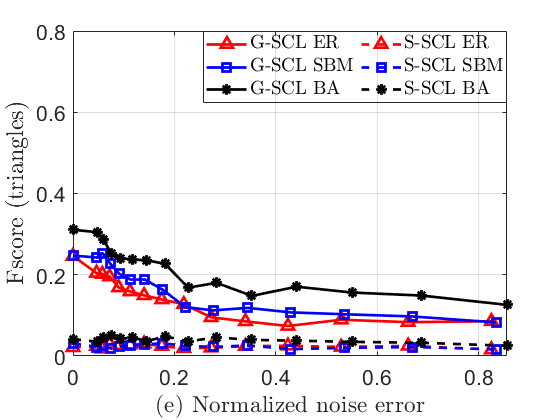}
    \end{subfigure}
    \begin{subfigure}[b]{0.302\textwidth}
        \includegraphics[width=\textwidth]{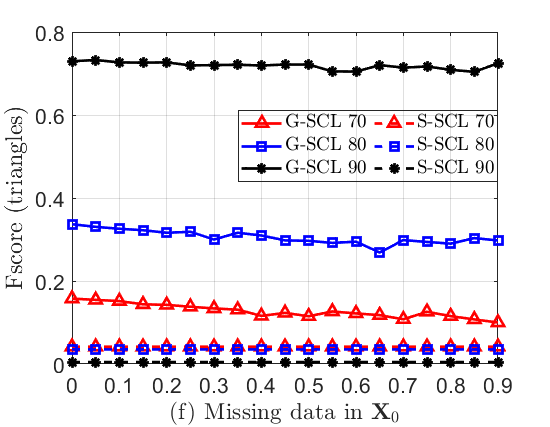}
    \end{subfigure}

    \caption{The first row shows the detection performance (F-score) for edges, while the second row displays the F-score for filled triangles. Each column evaluates the performance under different conditions: the number of samples (left column), the normalized noise level in the signals (middle column), and the proportion of missing data in $\bbX_0$ (right column).}
    \label{fig:topo_rec}
    \vspace{-0.5cm}

\end{figure*}

In this subsection, we evaluate the performance of the proposed method under different perturbations affecting the observed signals. Specifically, we consider limited sample size, additive noise, and partial observability, where a fraction of node signals is missing. For each scenario, multiple experimental configurations are used to assess robustness across diverse operating conditions.

\vspace{0.1cm}
\noindent\textbf{Limited data.}
We first explore three approaches to generate synthetic nodal data based on the eigenvalues of the graph Laplacian. Smooth node-level signals $\bbx \in \reals^N$ are generated by filtering white, zero-mean noise $\bbx_0 \in \reals^N$ using low-pass filters derived from the graph structure (here, ER graphs) encoded in the Laplacian matrix $\bbL_0 \in \reals^{N \times N}$: $\bbx = h(\bbL_0)\bbx_0$, where the choice of $h(\cdot)$ determines the filtering behavior. The considered filters are:
\noindent\textbf{Filter 1.} Low-pass filter based on Tikhonov regularization: $h(\lambda) = \frac{1}{1+\zeta\lambda}$, where larger $\zeta$ yields smoother signals.
\noindent\textbf{Filter 2.} Low-pass filter using the inverse eigenvalues of $\bbL$: $h(\lambda) = \frac{1}{\lambda}$.
\noindent\textbf{Filter 3.} Low-pass filter based on heat diffusion:
$h(\lambda) = \exp(-\zeta\lambda)$, with larger $\zeta$ again producing smoother signals.
Edge-level signals with low curl components are generated analogously by filtering with the upper Laplacian $\bbL_U = \bbB_2\bbB_2^\top$, which captures the structure of filled triangles \cite{isufi2022convolutional}.
Fig.~\ref{fig:topo_rec}(a) and (d) report the F-scores of the estimated edges and triangles, respectively, as the number of node and edge samples increases for G-SCL and S-SCL. G-SCL consistently achieves higher F-scores, reflecting more accurate recovery of the underlying simplicial structure. This improvement stems from jointly estimating edges and triangles using both nodal and edge signals, which better exploits the coupling between simplicial structures of order 1 and 2. Consequently, G-SCL requires fewer samples to achieve stable recovery, demonstrating higher sample efficiency. Regarding the signal-generation filters, Filter~3 (with $\zeta=1$) yields the smoothest signals and the best performance, while the Tikhonov filter \blue{(with $\zeta=1$)} produces less smooth signals and lower F-scores, as expected. Note that unobserved node signals are spread randomly over the graph and time (in contrast to unobserved edge signals), but they can be handled in the same way as the unobserved edges.

\vspace{0.1cm}
\noindent\textbf{Noisy data.} The second experiment evaluates robustness to additive noise in nodal and edge signals generated using Filter~3. The number of samples is fixed to $P_0 = P_1 = 10^3$, while increasing noise levels are added to both $\bbX_0$ and $\bbX_1$. Fig.~\ref{fig:topo_rec}(b) and (e) show the F-score for recovered edges and triangles as a function of noise intensity for three graph types. As expected, recovery performance degrades with increasing noise. However, G-SCL remains consistently more robust, exhibiting a slower performance decline than S-SCL, particularly for triangle estimation. This behavior is due to G-SCL’s ability to leverage complementary nodal and edge information, which provides additional constraints and mitigates noise effects. In contrast, S-SCL estimates edges and triangles independently, without coupling the two estimation tasks, resulting in higher sensitivity to noise, especially at high noise levels. \blue{Since triangle recovery depends on the correct estimation of the underlying edges, the poor edge recovery achieved by S-SCL under high noise levels propagates to the triangle estimation, leading to the near-zero triangle F-score observed in Fig.~\ref{fig:topo_rec}(e). A similar trend can also be observed for G-SCL at very high noise levels, although its degradation is less pronounced.} Across graph types, BA achieves the best edge and triangle recovery, while ER and SBM exhibit similar performance, likely due to structural differences in edge and triangle variability.

\vspace{0.1cm}
\noindent\textbf{Missing nodal data.} The third experiment examines the impact of missing node observations in a noise-free setting. Samples are generated using BA graphs and Filter~3, and missing data is introduced by randomly masking entries of $\bbX_0$. We consider three proportions of observed edge signals: {70\%, 80\%, 90\%}. As shown in Fig.~\ref{fig:topo_rec}(c) and (f), the recovery is not very sensitive to the fraction of missing node data but it is sensitive to the fraction of missing (available) edge data. Despite this, G-SCL consistently outperforms S-SCL. For the case of edge recovery [panel (c)], the gap grows as more nodal data is missing. Triangle recovery is harder for both methods, with S-SCL failing to infer triangles due to the large number of candidates relative to filled ones. In this regime, smoothness-based estimation becomes ill-conditioned. Higher edge observation ratios improve F-scores for both edges and triangles, as available edge signals partially compensate for missing node information.

\vspace{0.1cm}
\noindent\textbf{Ranking of wrongly estimated edges and triangles.} To complement Fig.~\ref{fig:topo_rec}, we analyze the smoothness rankings of wrongly estimated edges and triangles. For each element, we compute a smoothness measure and rank all candidates accordingly. The average ranking of misclassified elements reflects their proximity to satisfying the smoothness assumption. Results show that the proposed method not only improves edge and triangle recovery, but also assigns consistently lower (i.e., better) rankings to misclassified elements compared to competing methods. This indicates that incorrectly estimated edges and triangles are typically associated with smoother signals and are closer to being correctly recovered. Additional analyses in the project repository (see footnote 1) confirm these trends across all experimental configurations in Fig.~\ref{fig:topo_rec}.







 \begin{figure}[t]
    \centering

    \begin{subfigure}[b]{0.24\textwidth}
        \includegraphics[width=\textwidth]{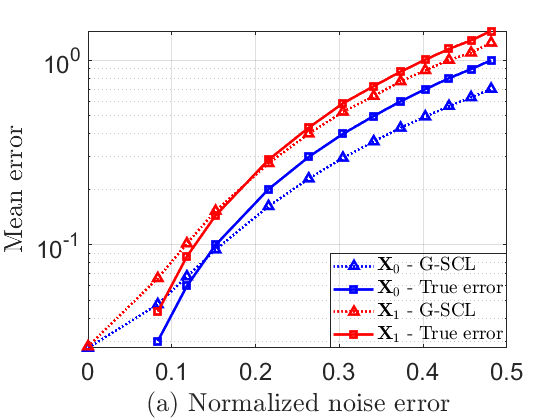}
    \end{subfigure}
    \begin{subfigure}[b]{0.24\textwidth}
        \includegraphics[width=\textwidth]{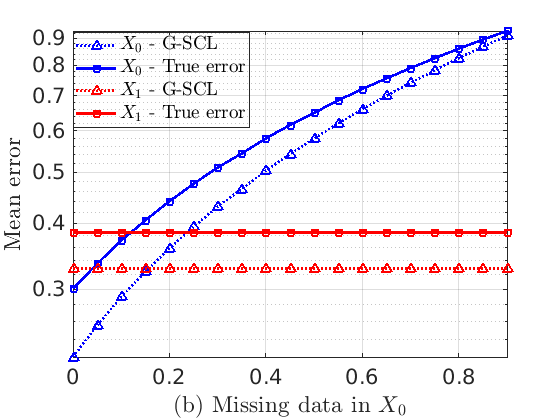}
    \end{subfigure}

    \caption{Denoising performance on $\bbX_0$ and $\bbX_1$ as a function of (a) normalized noise and (b) missing data ratio in $\bbX_0$.
}
    \label{fig:denoising}
\end{figure}

 \subsection{Signal  denoising/recovery}
 
In this second block, we evaluate the ability of our approach in denoising node and edge signals under different scenario across various perturbations in nodal signals.

\vspace{0.1cm}
\noindent\textbf{Denoising in $\bbX_0$ and $\bbX_1$.}
We generate $M=10^3$ nodes and edge signals considering ER graphs. The results presented in Fig.~\ref{fig:denoising}(a) show the normalized mean square error (NMSE) for $\bbX_0$ and $\bbX_1$ as we increase the noise level in the signals. 
In Fig.~\ref{fig:denoising}(a), we observe that for low noise levels, the denoising effect is minimal. As the noise increases, our method better approximates the ground truth, indicating improved denoising performance. This behavior suggests that when noise is low, the estimated topology does not contribute significantly to the denoising process, whereas for higher noise levels, the inferred structure helps filter out noise and recover a cleaner signal.
Additionally, denoising performance is consistently better for $\bbX_0$ than for $\bbX_1$, likely due to the lower availability of filled triangles in the estimated SC, which limits the structural support available for denoising $\bbX_1$.

\vspace{0.1cm}
\noindent\textbf{Missing data in $\bbX_0$.} The plot in Fig.~\ref{fig:denoising} (b) shows the denoising performance when fixing the noise level and varying the proportion of missing data in $\bbX_0$. As expected, the denoising performance for $\bbX_0$ degrades as the amount of missing data increases. This is attributed to a less accurate estimation of the graph and SC structure, which reduces the effectiveness of the denoising process. In contrast, denoising performance for $\bbX_1$ remains stable across different levels of missing data in $\bbX_0$, indicating that it is largely unaffected by this particular perturbation.

\begin{figure}[t]
    \centering
    
    \begin{subfigure}[b]{0.24\textwidth}
        \includegraphics[width=\textwidth]{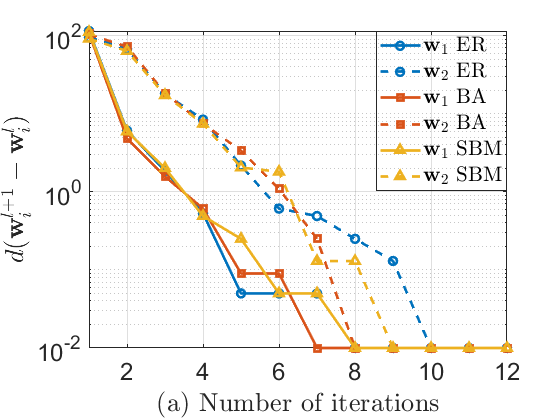}
    \end{subfigure}
    \begin{subfigure}[b]{0.24\textwidth}
        \includegraphics[width=\textwidth]{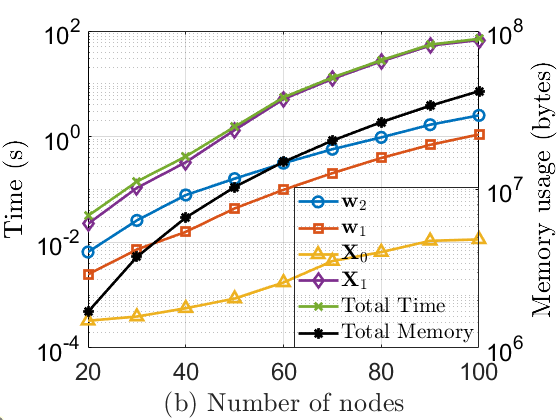}
    \end{subfigure}

    \caption{\blue{(a) Convergence behavior of the proposed approach in terms of the estimated edges and triangles. (b) Average computation time per iteration for each subproblem (left y-axis) and memory usage in bytes (right y-axis) versus the number of nodes (x-axis), using gradient descent to update $\bbX_0$ and $\bbX_1$.}}

    \label{fig:time_complexity_grad_convergence}
\end{figure}

\subsection{Algorithm Insights}
\vspace{0.1cm}
\noindent\textbf{Empirical convergence analysis.}  
We first study the number of external iterations required for convergence of the estimated edge and triangle structures. We consider graphs with $N = 20$ nodes and $M = 10^3$ node and edge signals. In each realization, $50\%$ of the edge signals are observed and $50\%$ of the triangles are assumed to be filled. Over 100 independent SC realizations, we measure the average difference between successive estimates of $\bbw_1$ (edges) and $\bbw_2$ (triangles).

As shown in Fig.~\ref{fig:time_complexity_grad_convergence}(a), the edge structure $\bbw_1$ converges faster, requiring approximately 7 external iterations, whereas the triangle structure $\bbw_2$ converges in about 9 iterations. This behavior is intuitive: since node signals are fully observed, edge estimation stabilizes earlier, while triangle estimation relies on partially observed edge signals and is therefore more challenging. We also note that two of the four subproblems admit closed-form solutions, while the remaining two are solved via greedy algorithms based on sorting operations. This combination enables efficient updates and explains the small number of iterations required for convergence. Finally, convergence behavior is consistent across ER, BA, and SBM graphs, with no significant differences observed.

\vspace{0.1cm}
\noindent\textbf{Empirical Time Complexity Analysis:}  
We next evaluate computational complexity as a function of the number of nodes by estimating SCs and their corresponding Hodge Laplacians of varying sizes. The number of samples is fixed to $M = 10^3$, and the average computation time per iteration for each subproblem is measured over 30 SC realizations per graph size. \blue{Fig.~\ref{fig:time_complexity_grad_convergence}(b) illustrates the average computation time per iteration for each subproblem (left y-axis) and the total memory usage in bytes (right y-axis) as a function of the number of nodes.}

The most computationally demanding subproblems correspond to the updates of $\bbw_2$ and $\bbX_1$, as both variables scale with the number of candidate edges, which is $\mathcal{O}(N^2)$. These updates involve matrix multiplications: for $\bbw_2$, in computing triangle smoothness scores; and for $\bbX_1$, during the gradient-based update. 
Additional analyses in the project repository consider closed-form updates for $\bbX_0$ and $\bbX_1$, showing that for graphs with more than 45 nodes, updating $\bbX_1$ becomes more computationally expensive than updating $\bbw_2$. This reversal reflects the increasing dimensionality of the matrices involved, where matrix inversion in the $\bbX_1$ update eventually dominates the cost of the multiplications required for updating $\bbw_2$. \blue{Furthermore, the right y-axis of Fig.~\ref{fig:time_complexity_grad_convergence}(b) demonstrates that the total memory required for the four estimated variables grows significantly with the network size. This escalating memory demand is directly tied to the number of edges in the graph, which dictates the dimensions of the stored variables, most notably $\bbX_1$, which is the largest matrix in the formulation.}

\blue{
\vspace{0.1cm}
\noindent\textbf{Effect of $\gamma$ under the relaxed SCC formulation.}
The SCC constraint is relaxed through the penalty term
$\gamma(\mathbf{1}-\bbw_1)^\top \bar{\bbB}_2^+ \bbw_2$. We evaluate the effect of $\gamma$ on the convergence of the SCC violation and the final reconstruction performance.
\begin{figure}[t]
    \centering

    \begin{subfigure}[b]{0.24\textwidth}
        \includegraphics[width=\textwidth]{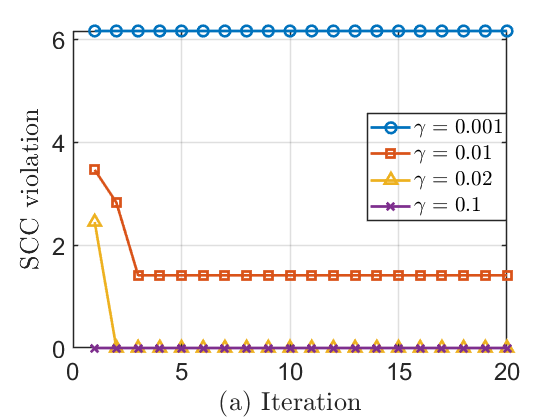}
    \end{subfigure}
    \hfill
    \begin{subfigure}[b]{0.24\textwidth}
        \includegraphics[width=\textwidth]{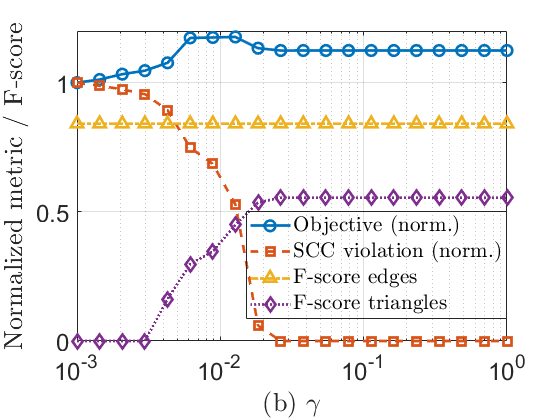}
    \end{subfigure}

    \caption{\blue{Effect of the SCC penalty parameter $\gamma$. (a): evolution of the pre-completion SCC violation $\|\bbB_1\bbB_2\|_F$ over the algorithm iterations for different values of $\gamma$. (b): normalized objective value, normalized SCC violation, and edge/triangle F-scores versus $\gamma$.}}
    \label{fig:gamma}
\end{figure}
Fig.~\ref{fig:gamma}(a) shows the evolution of the pre-completion SCC violation,
$\|\bbB_1\bbB_2\|_F$, over the algorithm iterations for a fixed graph realization.
For $\gamma=10^{-3}$ and $10^{-2}$, the violation decreases but converges to a nonzero value, requiring the final closure step to obtain a valid simplicial complex. In contrast, for $\gamma=2\!\cdot\!10^{-2}$ and $10^{-1}$, the violation reaches numerical zero after a few iterations, indicating that the optimization alone satisfies the SCC constraint.
The corresponding final performance is shown in Fig.~\ref{fig:gamma}(b). As $\gamma$ increases, the normalized SCC violation decreases monotonically and becomes zero beyond a threshold value. At the same time, the edge F-score remains nearly unchanged, while the triangle F-score improves until reaching a plateau. These results show that increasing $\gamma$ effectively promotes SCC-consistent solutions without degrading the reconstruction accuracy.
}

\begin{figure}[t]
    \centering

    \begin{subfigure}[b]{0.24\textwidth}
        \includegraphics[width=\textwidth]{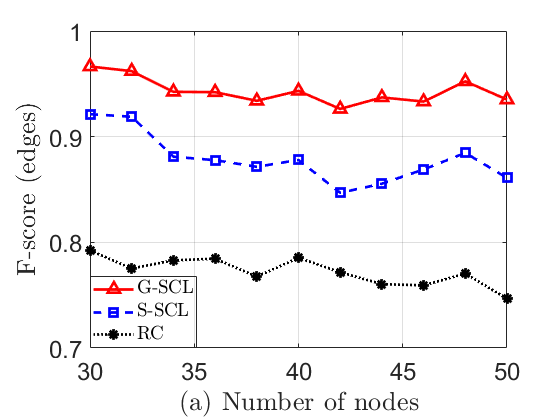}
    \end{subfigure}
        \begin{subfigure}[b]{0.24\textwidth}
        \includegraphics[width=\textwidth]{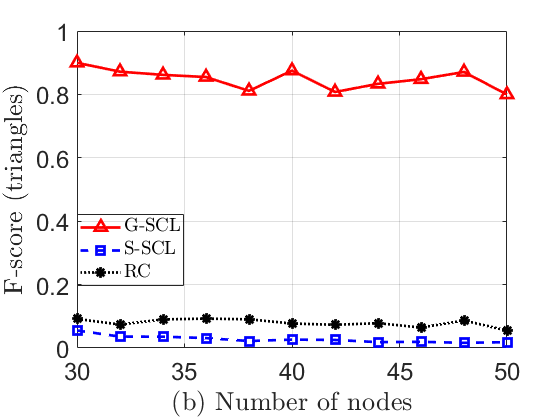}
    \end{subfigure}

    \caption{\blue{Edge (a) and triangle (b) F-score versus the number of nodes in a coauthorship dataset.}
}
    \label{fig:real_data}
\end{figure}

\subsection{\blue{Co-authorship dataset}}

\vspace{0.1cm}
We further evaluate the proposed approach on a real-world co-authorship dataset, comparing G-SCL with S-SCL and RC. \blue{Specifically, we employ the ACM dataset, originally compiled from the ACM Digital Library and preprocessed by \cite{wang2019heterogeneous}. This dataset has served as a benchmark for higher-order topology estimation in recent literature, including studies on learning simplicial complexes and hypergraphs from data \cite{buciulea2024learningicassp, tang2023learning}.} We construct \blue{the network from a subset of this dataset comprising papers published in top-tier conferences (KDD, SIGMOD, SIGCOMM, MobiCOMM, and VLDB). The resulting structure consists of authors as nodes, where} edges exist if two authors have co-authored at least one paper. Only coauthorship data is used to define the topology.
For signal generation of each author, we rely exclusively on the keywords associated with each paper. Node signals are created by aggregating the keyword vectors from all papers authored by each individual. Edge signals are similarly generated by aggregating the keyword vectors from papers co-authored by each pair of connected authors. 

Using these node and edge signals, we apply each method to estimate the SC. We consider a triangle as filled if three authors collaborate on the same paper, and assume that $70\%$ of the edge signals are observed.
Fig.~\ref{fig:real_data}(a) and Fig.~\ref{fig:real_data}(b) show the \blue{F-score} for edges and triangles, respectively, as the number of nodes in the graph increases. G-SCL consistently outperforms both S-SCL and RC in recovering edges and filled triangles. S-SCL performs comparably to G-SCL in edge recovery but fails to accurately estimate triangles due to its decoupled estimation process and limited ability to exploit edge information. Conversely, RC performs poorly in edge recovery but achieves better triangle estimation than S-SCL—though still below G-SCL. These results suggest the proposed approach can uncover aspects of higher-order structure when the signals conform to the assumed smoothness model.

\section{Conclusions}\label{Sec:Conclusions}
This paper presented a novel optimization-based framework for learning the topology of SCs from noisy and partially observed signals. A key contribution of our work was casting the learning problem as a rigorous high-dimensional and non-convex optimization that relates the topology of the SC with the observed signals. By modeling the SC topology using binary edge and triangle selection vectors and implementing a BCD algorithm, our approach ensures computational tractability while maintaining theoretical guarantees for convergence to a coordinate-wise minimum that is also a stationary point \blue{of the relaxed
penalized formulation}. Notably, our algorithm achieves modest computational complexity, a significant accomplishment given the challenge of inferring higher-order interactions. The proposed algorithm iteratively optimizes over blocks of variables, ensuring optimality within each block while progressively refining the overall solution. This design allowed us to incorporate sparsity and smoothness constraints on both node and edge signals, reflecting realistic data behaviors. Furthermore, we validated the effectiveness of our method through meaningful experiments on both synthetic and real-world datasets, demonstrating its ability to uncover higher-order relationships in complex data. Future work will extend this framework to consider SCs of higher orders, enabling the modeling of more intricate topological structures. Additionally, alternative approaches to relate signal properties to the SC topology will be explored, broadening the applicability and impact of SC-based learning methods.

\appendices 

\section{Proof of optimality of greedy allocation} \label{App:optimality of greedy algorithms}
This appendix presents the proofs of Lemmas~\ref{Lemma:GreedyTriangleSelection} and~\ref{Lemma:GreedyEdgeSelection}. As the two problems are very similar, we first prove Lemma~\ref{Lemma:GreedyTriangleSelection} and then briefly explain the modifications to prove Lemma~\ref{Lemma:GreedyEdgeSelection}.

\subsection{Proof of Lemma \ref{Lemma:GreedyTriangleSelection}}\label{App:ProofLemmaGreedyTriangleSelection}
The optimization problem w.r.t. $\bbw_2$ in \eqref{E:SC_OPT_v1_subp2_c1e} can be written as
\begin{alignat}{2}
	\!\!&\!\min_{\bbw_2\in\! \{0,1\}^{\bar{T}}} \alpha_2\|\bbw_2\|_0+ \beta_2\textstyle \sum_{t=1}^{\bar{T}}[\bar{\bbB}_2^\top \bar{\bbX}_1 \bar{\bbX}_1^\top\bar{\bbB}_2]_{tt}[\bbw_2]_t  	\label{E:SC_OPT_v1_subp2_c1e_app} \\ 
	\!\!&\! + \gamma \textstyle \sum_{t=1}^{\bar{T}}[\bar{\bbB}_2^{+\top}(\bbone-\bbw_1)]_{t}[\bbw_2]_t~\mathrm{\;\;s. \;t. } \,  \|\bbw_2\|_0 \!\geq \blue{T^{\text{budget}}}\! . \nonumber
\end{alignat} 
Since the entries of $\bbw_2$ are binary, we have that $\|\bbw_2\|_0=\sum_{t=1}^{\bar{T}}[\bbw_2]_t$. As a result, the optimization can be rewritten as
\begin{alignat}{2}
	\!\!&\!\min_{\bbw_2\in\! \{0,1\}^{\bar{T}}} \sum_{t=1}^{\bar{T}}\Big(\alpha_2[\bbw_2]_t +\beta_2[\bar{\bbB}_2^\top \bar{\bbX}_1 \bar{\bbX}_1^\top\bar{\bbB}_2]_{tt}[\bbw_2]_t   
	\label{E:SC_OPT_triangleselection_v2_app} \\ 
	\!\!&\!+\gamma [\bar{\bbB}_2^{+\top}(\bbone-\bbw_1)]_{t}[\bbw_2]_t\Big) ~\mathrm{\;\;s. \;t. } \,  \sum_{t=1}^{\bar{T}}[\bbw_2]_t \!\geq \blue{T^{\text{budget}}}\! . \nonumber
\end{alignat} 
Upon defining the triangle-score vector (cf. Lemma  \ref{Lemma:GreedyTriangleSelection})
$$[\bbs_2]_t=\alpha_2+\beta_2[\bar{\bbB}_2^\top \bar{\bbX}_1 \bar{\bbX}_1^\top\bar{\bbB}_2]_{tt} + \gamma [\bar{\bbB}_2^{+\top}(\bbone-\bbw_1)]_{t}$$
for $t=1,...,\bar{T}$, the optimization is simply
\begin{alignat}{2}
	\!\!&\!\min_{\bbw_2\in\! \{0,1\}^{\bar{T}}} \sum_{t=1}^{\bar{T}}[\bbs_2]_t[\bbw_2]_t
 ~\mathrm{\;\;s. \;t. } \,  \sum_{t=1}^{\bar{T}}[\bbw_2]_t\!\geq \blue{T^{\text{budget}}}\!.	\label{E:SC_OPT_triangleselection_v3_app} 
\end{alignat} 
Since all entries in $\bbs_2$ are positive, any entry of $\bbw_2$ that takes the value 1 increases the cost in \eqref{E:SC_OPT_triangleselection_v3_app}. On the other hand, the constraint in \eqref{E:SC_OPT_triangleselection_v3_app} requires at least $\blue{T^{\text{budget}}}$ of those entries to be one. As a result, the optimal solution is to sort vector $\bbs_2$ increasingly, take the (first) $\blue{T^{\text{budget}}}$ indexes associated with the smallest values of the score, set the entries of $\bbw_2$ associated with these $\blue{T^{\text{budget}}}$ indexes to 1, and set all other entries of $\bbw_2$ to zero. This is precisely the allocation proposed in Lemma \ref{Lemma:GreedyTriangleSelection}.

\subsection{Proof of Lemma \ref{Lemma:GreedyEdgeSelection}}\label{App:ProofLemmaGreedyEdgeSelection}
The proof is analogous to the one for Lemma \ref{Lemma:GreedyTriangleSelection}, with the main difference being that if $l\in \ccalEobs$, then we need to guarantee that $[\bbw_1]_l=1$. 
Specifically, the optimization problem w.r.t. $\bbw_1$ in \eqref{E:SC_OPT_v1_subp1_obj} can be rewritten as
\begin{align}\label{E:SC_OPT_edgeselection_v2_app}
		\!\!&\!\min_{\bbw_1\in\! \{0,1\}^{\bar{E}}} \sum_{l=1}^{\bar{E}}\Big(\alpha_1[\bbw_1]_l +\beta_1[\bar{\bbB}_1^\top\blue{\bar{\bbX}_0\bar{\bbX}_0^\top}\bar{\bbB}_1]_{ll} [\bbw_1]_l   \\ 
	\!\!&\!- \gamma [\bar{\bbB}_2^+\bbw_2]_{l}[\bbw_1]_l\Big) ~\!\!\mathrm{\;\;s. \;t. } ~ [\bbw_1]_{l}\!=\!1 \;\forall \;l\!\in\! \ccalEobs\!, \sum_{l=1}^{\bar{E}}[\bbw_1]_l \!\geq \!E^{\min}\!. \nonumber 
\end{align} 
Upon defining the edge-score vector (cf. Lemma  \ref{Lemma:GreedyEdgeSelection})
$$[\bbs_1]_l=\alpha_1+\beta_1[\bar{\bbB}_1^\top\blue{\bar{\bbX}_0\bar{\bbX}_0^\top}\bar{\bbB}_1]_{ll} - \gamma [\bar{\bbB}_2^+\bbw_2]_{l}$$
for $l\notin \ccalEobs$ and $[\bbs_1]_l=-1$ for $l\in \ccalEobs$, the optimization can be rewritten as
\begin{alignat}{2}
	\!\!&\!\min_{\bbw_1\in\! \{0,1\}^{\bar{E}}} \sum_{l=1}^{\bar{E}}[\bbs_1]_l[\bbw_1]_l
 ~\mathrm{\;\;s. \;t. } \,  \sum_{l=1}^{\bar{E}}[\bbw_1]_l\!\geq E^{\min}\!.	\label{E:SC_OPT_edgeselection_v3_app} 
\end{alignat} 
As in \eqref{E:SC_OPT_triangleselection_v3_app}, the optimal solution is to activate the edges with the smallest score. If there are more than $E^{\min}$ entries in $\bbs_1$ with a negative score  (i.e., if the parameter $E^{\mathrm{neg}}$ defined in the lemma is greater or equal than $E^{\min}$), then all of the associated edges are activated. On the other hand, if the $E^{\min}$-th smallest score is positive, then the edge cardinality constraint in \eqref{E:SC_OPT_edgeselection_v3_app}  is active and the optimal solution activates the edges associated with the $E^{\min}$ smallest scores. Note that, all the links in $\ccalEobs$ are activated since the value of the score for those links is negative. This is the allocation proposed in Lemma \ref{Lemma:GreedyEdgeSelection}.

\section{Proof of Theorem \ref{Theo:ConvergenceBlockCoordinate} (convergence of  BCD)} \label{App:ProofTheoConvergenceBlockCoordinate}
The proof is based on the convergence of the BCD algorithm \cite{tseng2001convergence} for nondifferentiable minimization.
In summary, \cite{tseng2001convergence} analyzes the convergence of BCD for problems of the form
\begin{eqnarray}\label{eq:problem_formul_cong_BCD_smooth_plus_nonsmooth}
&f(x) \;=\; f_0(x_1,\dots,x_N) + \sum_{k=1}^N f_k(x_k),&
\end{eqnarray}
where $f_0$ is a differentiable (smooth) coupling term and each $f_k$ is a (possibly nondifferentiable) function of block $x_k$ only. 
To ensure that any limit point of the BCD iterates is a stationary point of $f$, \cite{tseng2001convergence} shows the following conditions must hold:
Then, \cite[Th. 4]{tseng2001convergence} proves that BCD algorithms converge to a stationary point when 4 conditions are fulfilled (see two first paragraphs of App. \ref{App:SubappConvergenceBCD} for a detailed description of the conditions).

The results in \cite{tseng2001convergence} apply only to continuous optimization. To use them for \eqref{E:SC_OPT_v1}, which includes binary variables, we proceed in three steps: s1) we consider the continuous relaxation \eqref{E:SC_OPT_v1_relaxed}; s2) we solve \eqref{E:SC_OPT_v1_relaxed} via alternating minimization and show that the solutions in Lemmas \ref{Lemma:GreedyTriangleSelection}–\ref{Lemma:EdgeSignalsInterp}, originally derived for \eqref{E:SC_OPT_v1}, also solve its relaxed subproblems; and s3) we show that \eqref{E:SC_OPT_v1_relaxed} satisfies the conditions of \cite[Th.~4]{tseng2001convergence}, implying \eqref{E:convergenceTheorem} in Th. \ref{Theo:ConvergenceBlockCoordinate}. Details of the steps s2) and s3) are given in App. \ref{App:SubappEquivalenceRelaxBinary} and \ref{App:SubappConvergenceBCD}.

\subsection{Optimality of Lemmas \ref{Lemma:GreedyTriangleSelection}-\ref{Lemma:EdgeSignalsInterp} for the relaxed formulation}\label{App:SubappEquivalenceRelaxBinary}
Using the results in App. \ref{App:optimality of greedy algorithms}, the optimization in \eqref{E:SC_OPT_v1_relaxed} w.r.t. each of the four blocks of variables can be written as:
\begin{align}\label{E:SC_OPT_edgeselection_v2_app_relaxed}
		\!\!&\!\min_{\bbw_1\in\! [0,1]^{\bar{E}}} \sum_{l=1}^{\bar{E}}\Big(\alpha_1[\bbw_1]_l +\beta_1[\bar{\bbB}_1^\top\blue{\bar{\bbX}_0\bar{\bbX}_0^\top}\bar{\bbB}_1]_{ll} [\bbw_1]_l   \\ 
	\!\!&\!- \gamma [\bar{\bbB}_2^+\bbw_2]_{l}[\bbw_1]_l\Big) ~\!\!\mathrm{\;\;s. \;t. } ~ [\bbw_1]_{l}\!=\!1 \;\forall \;l\!\in\! \ccalEobs\!, \sum_{l=1}^{\bar{E}}[\bbw_1]_l \!\geq \!E^{\min}\!; \nonumber \\
\!\!&\!\min_{\bbw_2\in\! [0,1]^{\bar{T}}} \sum_{t=1}^{\bar{T}}\Big(\alpha_2[\bbw_2]_t +\beta_2[\bar{\bbB}_2^\top \bar{\bbX}_1 \bar{\bbX}_1^\top\bar{\bbB}_2]_{tt}[\bbw_2]_t   
	\label{E:SC_OPT_triangleselection_v2_app_relaxed} \\ 
	\!\!&\!+\gamma [\bar{\bbB}_2^{+\top}(\bbone-\bbw_1)]_{t}[\bbw_2]_t\Big) ~\mathrm{\;\;s. \;t. } \,  \sum_{t=1}^{\bar{T}}[\bbw_2]_t \!\geq \blue{T^{\text{budget}}}\!; \nonumber\\
	\!\!&\!\min_{\bar{\bbX}_{0}}  \; \beta_1\tr(\bar{\bbX}_0 \bar{\bbX}_0^\top\bar{\bbB}_1\diag(\bbw_1)\bar{\bbB}_1^\top) + \eta_0\| \bar{\bbX}_0-\bbX_0^\ccalO\|_F^2; \label{E:SC_OPT_v1_subp3_obj_app_relaxed} \\ 
	\!\!\!&\min_{\bar{\bbX}_{1}} \; \beta_2\tr(\bar{\bbX}_1 \bar{\bbX}_1^\top\bar{\bbB}_2\diag(\bbw_2)\bar{\bbB}_2^\top) \nonumber\\
    & \hspace{3cm}+ \eta_1\|\bar{\bbTheta} \bar{\bbX}_1-\bbX_1^\ccalO\|_F^2+\varepsilon\|\bar{\bbX}_1\|_F^2.  \label{E:SC_OPT_v1_subp4_obj_app_relaxed} 
\end{align} 
The problems \eqref{E:SC_OPT_v1_subp3_obj_app_relaxed}  and  \eqref{E:SC_OPT_v1_subp4_obj_app_relaxed} 
 are exactly the same as those considered in \eqref{E:SC_OPT_v1_subp3_obj}  and  \eqref{E:SC_OPT_v1_subp4_obj}, so the solutions provided in Lemmas \ref{Lemma:NodalSignalsDenoised} and \ref{Lemma:EdgeSignalsInterp} are optimal here as well.  

Regarding problems \eqref{E:SC_OPT_edgeselection_v2_app_relaxed}  and  \eqref{E:SC_OPT_triangleselection_v2_app_relaxed}, we observe that the only difference with those in \eqref{E:SC_OPT_edgeselection_v2_app}  and  \eqref{E:SC_OPT_triangleselection_v2_app} (which are equivalent to the versions presented in Section \ref{S:GreedyAlgorithmAlgorithm}) is that the binary domain constraints $\{0,1\}\times...\times\{0,1\} $ have been replaced with their convex counterparts $[0,1]\times...\times[0,1] $. Interestingly, the relaxed problems are linear and separable across the entries of the selection vectors, with the only coupling being given by the cardinality constraints. As a result: i) the optimal solution sets each entry either to its maximum value (one) or its minimum value (zero), ii) the entries that are set to one are those with the smallest score, and iii) the number of ones is either the number of negative entries or the minimum value set by the cardinality constraint, whichever is larger. It is easy to see that the solutions in Lemmas~\ref{Lemma:GreedyTriangleSelection}~and~\ref{Lemma:GreedyEdgeSelection} satisfy these conditions and, therefore, are also optimal for \eqref{E:SC_OPT_edgeselection_v2_app_relaxed}  and  \eqref{E:SC_OPT_triangleselection_v2_app_relaxed}.

\subsection{Convergence of the relaxed formulation using BCD}\label{App:SubappConvergenceBCD} We start by listing the conditions presented in \cite{tseng2001convergence} that must hold in order \blue{for} the BCD algorithm to converge to a stationary point. Considering BCD problems of the form 
in \eqref{eq:problem_formul_cong_BCD_smooth_plus_nonsmooth}, where we recall that $f_0$ is a differentiable (smooth) coupling term and each $f_k$ is a (possibly nondifferentiable) function of block $x_k$ only. To ensure that any limit point of the BCD iterates is a stationary point of $f$, \cite{tseng2001convergence} provides several sets of sufficient conditions. We use the variant that requires:
\begin{itemize} 
\item \textit{Continuity and bounded level set:} $f$ must be continuous on the initial level set $\ccalX^0=\{\bbx:f(\bbx)\le f(\bbx^0)\}$, and $\ccalX^0$ must be compact. This guarantees that each block subproblem has an attainable minimizer. 
\item \textit{Block updates (cyclic rule):} The BCD algorithm must update blocks in an essentially cyclic manner, ensuring each block is updated infinitely often. 
\item \textit{Differentiability (regularity) of $f_0$:} The smooth coupling part $f_0$ must be continuously differentiable on an open domain, which implies the regularity condition in \cite{tseng2001convergence} is \blue{satisfied} for $f$, i.e., any coordinate-wise minimum is also a stationary point. 
\item \textit{Uniqueness:} Each block subproblem has a unique minimizer, so that limit points of BCD are stationary. \end{itemize} 
In summary, the result in \cite{tseng2001convergence} requires: (i) continuity with compact level sets, (ii) cyclic block updates, (iii) smoothness of $f_0$, and (iv) uniqueness of block solutions. Under these assumptions, any limit point of the BCD iterations is a stationary point of $f$. We now verify these conditions for the given problem in \eqref{E:SC_OPT_v1_relaxed}. The decision variables are $\bbw_1,\bbw_2$ (vectors) and $\bar\bbX_0,\bar\bbX_1$ (matrices). The objective can be written as 
\begin{align*} f(\bbw_1,\bbw_2,\bar\bbX_0,\bar\bbX_1) &= f_0(\bbw_1,\bbw_2,\bar\bbX_0,\bar\bbX_1) + f_1(\bbw_1) + f_2(\bbw_2).
\end{align*} 
The smooth part $f_0$ is
\begin{align*} 
f_0(\bbw_1,&\bbw_2,\bar\bbX_0,\bar\bbX_1) \! = \gamma(\mathbf{1}-\bbw_1)^\top \bar\bbB_2^+ \bbw_2\!+\!\varepsilon\| \bar{\bbX}_1\|_F^2 \\
&+\! \eta_0\|\bar\bbX_0-\bbX_0^\ccalO\|_F^2 \!+\! \beta_1\tr(\bar\bbX_0\bar\bbX_0^\top\bar\bbB_1\diag(\bbw_1)\bar\bbB_1^\top) \\ 
&+\! \eta_1\|\bar{\bbTheta}\bar\bbX_1-\bbX_1^\ccalO\|_F^2 \!+\! \beta_2\tr(\bar\bbX_1\bar\bbX_1^\top\bar\bbB_2\diag(\bbw_2)\bar\bbB_2^\top), \end{align*} 
which is differentiable (polynomial/quadratic) in all variables $(\bbw_1,\bbw_2,\bar\bbX_0,\bar\bbX_1)$. The remaining nonsmooth parts are \begin{align*} 
&f_1(\bbw_1) = \alpha_1\|\bbw_1\|_1 + I_{\{\bbw_1 \!\in\! [0,1]^{\bar{E}}, \; \sum_{l=1}^{\bar{E}} [\bbw_1]_l \!\geq\! E^{\min}, \; [\bbw_1]_l\!=\!1 \; \forall l\in\ccalEobs\}}, \hspace{.7cm}
\\ &f_2(\bbw_2) = \alpha_2\|\bbw_2\|_1 + I_{\{\bbw_2 \in [0,1]^{\bar{T}}, \; \sum_{t=1}^{\bar{T}} [w_2]_t \geq \blue{T^{\text{budget}}}\}}, 
\end{align*} 
where $I_{\{\cdot\}}$ are indicator functions enforcing the box and cardinality constraints. 

\emph{Compact level sets and continuity:} 
The objective $f$ is the sum of a continuously differentiable part $f_0$ and proper lower semicontinuous convex terms (the $\ell_1$ terms and indicator functions). Hence $f$ is lower semicontinuous. The feasible set induced by the box constraints $\bbw_i\in[0,1]^{\cdot}$ and the linear equalities/inequalities is closed and the $w$--components are therefore bounded. It remains to show the matrix variables $\bar\bbX_0,\bar\bbX_1$ cannot escape to infinity while keeping $f$ below a fixed level. This is ensured by the quadratic penalty terms $\eta_0\|\bar\bbX_0-\bbX_0^\ccalO\|_F^2$ and $\varepsilon\|\bar\bbX_1\|_F^2$: since $\eta_0>0,\varepsilon>0$ these terms are coercive in $\bar\bbX_0,\bar\bbX_1$ and thus $f\to\infty$ whenever $\|\bar\bbX_i\|\to\infty$. Therefore any sublevel set $\{x : f(x)\le f(x^0)\}$ is closed and bounded, hence compact in the finite-dimensional setting; in particular each block subproblem attains a minimizer.

\emph{Cyclic block updates:}
The BCD implementation updates the four blocks
$(\bbw_1 \;\rightarrow\; \bbw_2 \;\rightarrow\; \bar{\bbX}_0 \;\rightarrow\; \bar{\bbX}_1)$ in a fixed round-robin order.  
Hence each block is updated once per iteration cycle and therefore infinitely 
often.  
This satisfies the “essentially cyclic’’ update rule required by 
\cite{tseng2001convergence}.

\emph{Regularity and differentiability:} 
The coupling part $f_0$ is polynomial/quadratic and therefore continuously differentiable on the whole Euclidean space. The nonsmooth parts $f_1,f_2$ are proper, convex and lower semicontinuous. By Lemma 3.1 of \cite{tseng2001convergence}, these conditions imply that $f$ is regular (a coordinate-wise minimum is a stationary point). Thus the differentiability/regularity requirements for Tseng's result hold.

\emph{Uniqueness:} Our objective is not globally convex in all blocks (due to the bilinear terms $\bar\bbX\bar\bbX^\top \diag(\bbw)$, etc.). However, when optimizing over a single block with the others fixed, the coupling terms become linear (for $\bbw_i$) or quadratic (for $\bbX_i$), giving rise to convex per-block formulations that (under mild assumptions) have a unique minimizer. 
\begin{itemize} 
\item \textbf{$X$‐blocks ($\bar\bbX_0,\bar\bbX_1$):} If $\bbw_1,\bbw_2$ are fixed, the terms involving $\bar\bbX_0$ or $\bar\bbX_1$ are quadratic (squared Frobenius norms) plus positive semidefinite quadratic forms [cf. \eqref{E:SC_OPT_v1_subp3_obj_app_relaxed} and \eqref{E:SC_OPT_v1_subp4_obj_app_relaxed} ]. The Hessian of \eqref{E:SC_OPT_v1_subp3_obj_app_relaxed} w.r.t.\ $\bar\bbX_0$ is $2\eta_0 \bbI+2\beta_1\bar\bbB_1\diag(\bbw_1)\bar\bbB_1^\top$, which is positive definite since $\eta_0>0$. Hence the $\bar\bbX_0$‐subproblem is strictly convex and has a unique minimizer. The same holds for the Hessian of \eqref{E:SC_OPT_v1_subp4_obj_app_relaxed}  w.r.t. $\bar\bbX_1$, since $\varepsilon>0$. 
\item \textbf{$w$‐blocks ($\bbw_1,\bbw_2$):} With $\bar\bbX_0,\bar\bbX_1$ fixed, the optimization w.r.t. $\bbw_1$ is given in \eqref{E:SC_OPT_edgeselection_v3_app}, with $\bbw_1\in\! [0,1]^{\bar{E}}$ in lieu of $\bbw_1\in\! \{0,1\}^{\bar{E}}$. The objective and constraints are linear and separable across the entries of $\bbw_1$. The same holds true for the optimization w.r.t. $\bbw_2$ in \eqref{E:SC_OPT_triangleselection_v3_app}, after linearizing the binary constraints. \blue{Both problems are linear (convex) and separable across entries; hence, multiple optimal solutions can arise only from a tie at the active cardinality cut or from a free entry whose score is exactly zero (such an entry could take any value in $[0,1]$ without altering the objective). Both degeneracies are excluded under (AS5).} If each $w$‐block subproblem has a unique optimal solution (cf. Lemmas \ref{Lemma:GreedyTriangleSelection} and \ref{Lemma:GreedyEdgeSelection}), then \cite[Th. 4.1.(c)]{tseng2001convergence}  applies: every cluster point is a coordinate-wise minimum, and by regularity (from above) it is a stationary point. 
\end{itemize} 

\emph{Summary:} Under the \blue{no-degeneracy} assumption (AS5),  our problem in \eqref{E:SC_OPT_v1_relaxed}: i) yields a solution that is feasible for the penalized version of \eqref{E:SC_OPT_v1}; and ii)  satisfies the requirements in \cite{tseng2001convergence} and, thus, any limit point of the BCD iterates generated by Alg. \ref{Alg:GreedySCL} is  a stationary point of the penalized version of  \eqref{E:SC_OPT_v1_relaxed}.

\bibliographystyle{IEEEtran.bst}
\bibliography{citations}

@string{TSP = "IEEE Trans. Signal Process."}

@string{TIT = "IEEE Trans. Inf. Theory"}

@string{JSTSP = "IEEE J. Sel. Topics Signal Process."}

@string{SPMag = "IEEE Signal Process. Mag."}

@string{ICASSP = "IEEE Int. Conf. Acoust., Speech, Signal Process. (ICASSP)"}

@string{EUSIPCO = "Eur. Signal Process. Conf. (EUSIPCO)"}

@string{SAM ="IEEE Intl. Wrksp. Sensor Array Multichannel Signal Process. (SAM)"}

@string{ASILOMAR ="Asilomar Conf. Signals, Syst., Comput."}

@string{TSIPN ="IEEE Trans. Signal Inf. Process. Netw"}

@string{ICLR ="Intl. Conf. on Learning Representations (ICLR)"}

@inproceedings{chepuri2017learning,
  title={Learning sparse graphs under smoothness prior},
  author={Chepuri, S. and Liu, S. and Leus, G. and Hero, A. O.},
  booktitle=ICASSP,
  pages={6508--6512},
  year={2017},
  organization={IEEE}
}

@article{dong2016learning,
  title={Learning Laplacian matrix in smooth graph signal representations},
  author={Dong, X. and Thanou, D. and Frossard, P. and Vandergheynst, P.},
  journal=TSP,
  volume={64},
  number={23},
  pages={6160--6173},
  year={2016},
  publisher={IEEE}
}

@article{isufi2025topological,
  title={Topological signal processing and learning: Recent advances and future challenges},
  author={Isufi, E. and Leus, G. and Beferull-Lozano, B. and Barbarossa, S. and Di Lorenzo, P.},
  journal={Signal Processing},
  pages={109930},
  year={2025},
  publisher={Elsevier}
}

@article{saboksayr2021online,
  title={Online discriminative graph learning from multi-class smooth signals},
  author={Saboksayr, S. S. and Mateos, G. and Cetin, M.},
  journal={Signal Processing},
  volume={186},
  pages={108101},
  year={2021},
  publisher={Elsevier}
}

@article{buciulea2025polynomial,
  title={Polynomial graphical lasso: Learning edges from {G}aussian graph-stationary signals},
  author={Buciulea, A. and Ying, J. and Marques, A. G. and Palomar, D. P.},
  journal=TSP,
  year={2025},
  volume={73},
  pages={1153-1167}
}

@article{shuman2013emerging,
  author    = {Shuman, D. I. and Narang, S. K. and Frossard, P. and Ortega, A. and Vandergheynst, P.},
  title     = {The emerging field of signal processing on graphs: Extending high-dimensional data analysis to networks and other irregular domains},
  journal   = SPMag,
  volume    = {30},
  number    = {3},
  pages     = {83--98},
  year      = {2013}
}

@article{battiston2021physics,
  author    = {Battiston, F. and Amico, E. and Barrat, A. and Bianconi, G. and et al.},
  title     = {The physics of higher-order interactions in complex systems},
  journal   = {Nature Physics},
  volume    = {17},
  number    = {10},
  pages     = {1093--1098},
  year      = {2021}
}

@article{giusti2016twos,
  author    = {Giusti, C. and Ghrist, R. and Bassett, D. S.},
  title     = {Two's company, three (or more) is a simplex: Algebraic-topological tools for understanding higher-order structure in neural data},
  journal   = {J. Comput. Neuroscience},
  volume    = {41},
  number    = {1},
  pages     = {1--14},
  year      = {2016}
}

@article{patania2017shape,
  author    = {Patania, A. and Petri, G. and Vaccarino, F.},
  title     = {The shape of collaborations},
  journal   = {EPJ Data Sci.},
  volume    = {6},
  number    = {1},
  pages     = {18},
  year      = {2017}
}

@article{benson2018simplicial,
  author    = {Benson, A.R. and Abebe, R. and Schaub, M. T. and Jadbabaie, A. and Kleinberg, J.},
  title     = {Simplicial closure and higher-order link prediction},
  journal   = {Proc. Natl. Acad. Sci. U.S.A.},
  volume    = {115},
  number    = {48},
  pages     = {E11221--E11230},
  year      = {2018}
}

@article{egilmez2017graph,
  author    = {Egilmez, H. E. and Pavez, E. and Ortega, A.},
  title     = {Graph learning from data under Laplacian and structural constraints},
  journal   = JSTSP,
  volume    = {11},
  number    = {6},
  pages     = {1005--1016},
  year      = {2017}
}

@inproceedings{kalofolias2016learn,
  author    = {Kalofolias, V.},
  title     = {How to learn a graph from smooth signals},
  booktitle = {Int. Conf. Artif. Intell. Stat. (AISTATS)},
  pages     = {920--929},
  year      = {2016}
}

@article{thanou2017learning,
  author    = {Thanou, D. and Dong, X. and Kressner, D. and Frossard, P.},
  title     = {Learning heat diffusion graphs},
  journal   = TSIPN,
  volume    = {3},
  number    = {3},
  pages     = {484--499},
  year      = {2017}
}

@article{dong2019learning,
  author    = {Dong, X. and Thanou, D. and Rabbat, M. and Frossard, P.},
  title     = {Learning graphs from data: A signal representation perspective},
  journal   = SPMag,
  volume    = {36},
  number    = {3},
  pages     = {44--63},
  year      = {2019}
}

@inproceedings{tang2023hypergraphs,
  author    = {Tang, B. and Chen, S. and Dong, X.},
  title     = {Learning hypergraphs from signals with dual smoothness prior},
  booktitle = ICASSP,
  pages     = {1--5},
  year      = {2023}
}

@article{delabays2025hypergraph,
  author    = {Delabays, R. and De Pasquale, G. and D\"orfler, F. and Zhang, Y.},
  title     = {Hypergraph reconstruction from dynamics},
  journal   = {Nat. Commun.},
  volume    = {16},
  pages     = {2691},
  year      = {2025}
}

@article{nguyen2021learning,
  author    = {Nguyen, C.  H. and Mamitsuka, H.},
  title     = {Learning on hypergraphs with sparsity},
  journal   = {IEEE Trans. Pattern Anal. Mach. Intell.},
  volume    = {43},
  number    = {8},
  pages     = {2710--2722},
  year      = {2021}
}

@inproceedings{buciulea2024learningsam,
  title={Learning the topology of a simplicial complex using simplicial signals: A greedy approach},
  author={Buciulea, A. and Isufi, E. and Leus, G. and Marques, A. G.},
  booktitle=SAM,
  year={2024},
  organization={IEEE}
}

@article{tseng2001convergence,
  title={Convergence of a block coordinate descent method for nondifferentiable minimization},
  author={Tseng, P.},
  journal={Journal of optimization theory and applications},
  volume={109},
  pages={475--494},
  year={2001},
  publisher={Springer}
}

@article{gorski2007biconvex,
  title={Biconvex sets and optimization with biconvex functions: a survey and extensions},
  author={Gorski, J. and Pfeuffer, F. and Klamroth, K.},
  journal={Mathematical methods of operations research},
  volume={66},
  pages={373--407},
  year={2007},
  publisher={Springer}
}

@article{schaub2021SPOnHigherOrder,
title = {Signal processing on higher-order networks: Livin’ on the edge... and beyond},
journal = {Signal Processing},
volume = {187},
pages = {108149},
year = {2021},
issn = {0165-1684},
doi = {https://doi.org/10.1016/j.sigpro.2021.108149},
author = {M. T. Schaub and Y. Zhu and J.-B. Seby and T. M. Roddenberry and S. Segarra}
}

@article{mateos2019connecting,
  title={Connecting the dots: Identifying network structure via graph signal processing},
  author={Mateos, G. and Segarra, S. and Marques, A. G. and Ribeiro, A.},
  journal=SPMag,
  volume={36},
  number={3},
  pages={16--43},
  year={2019},
  publisher={IEEE}
}

@ARTICLE{segarra2017network, 
author={S. Segarra and A. G. Marques and G. Mateos and A. Ribeiro}, 
journal=TSIPN, 
title={Network Topology Inference from Spectral Templates}, 
year={2017}, 
volume={3}, 
number={3}, 
pages={467-483}, 
month={Sep.},}

@article{zomorodian2010fast,
  title={Fast construction of the {Vietoris-Rips} complex},
  author={A. Zomorodian },
  journal={Computers \& Graphics},
  volume={34},
  number={3},
  pages={263--271},
  year={2010},
  publisher={Elsevier}
}

@article{friedman2008sparse,
  title={Sparse inverse covariance estimation with the graphical lasso},
  author={Friedman, J. and Hastie, T. and Tibshirani, R.},
  journal={Biostatistics},
  volume={9},
  number={3},
  pages={432--441},
  year={2008},
  publisher={Oxford Univ. Press}
}

@article{timme2007revealing,
  title={Revealing network connectivity from response dynamics},
  author={Timme, M.},
  journal={Physical Review Letters},
  volume={98},
  number={22},
  pages={224101},
  year={2007},
  publisher={APS}
}

@article{barbarossa2020topological,
  title={Topological signal processing over simplicial complexes},
  author={Barbarossa, S. and Sardellitti, S.},
  journal=tsp,
  volume={68},
  pages={2992--3007},
  year={2020},
  publisher={IEEE}
}

@inproceedings{buciulea2024learningicassp,
  title={Learning graphs and simplicial complexes from data},
  author={Buciulea, A. and Isufi, E. and Leus, G. and Marques, A. G.},
  booktitle=icassp,
  pages={9861--9865},
  year={2024},
  organization={IEEE}
}

@article{wang2022full,
  title={Full reconstruction of simplicial complexes from binary contagion and {I}sing data},
  author={Wang, H. and Ma, C. and Chen, H.-S. and Lai, Y.-C. and Zhang, H.-F.},
  journal={Nature Comms.},
  volume={13},
  number={1},
  pages={3043},
  year={2022},
  publisher={Nature Publishing Group UK London}
}

@article{yang2022simplicial,
  title={Simplicial convolutional filters},
  author={Yang, M. and Isufi, E. and Schaub, M. T. and Leus, G.},
  journal=tsp,
  volume={70},
  pages={4633--4648},
  year={2022},
  publisher={IEEE}
}

@inproceedings{tang2023learning,
  title={Learning Hypergraphs From Signals With Dual Smoothness Prior},
  author={Tang, B. and Chen, S. and Dong, X.},
  booktitle=icassp,
  pages={1--5},
  year={2023},
  organization={IEEE}
}

@inproceedings{gurugubelli2024simplicial,
  title={Simplicial complex learning from edge flows via sparse clique sampling},
  author={Gurugubelli, S. and Chepuri, S. P.},
  booktitle=EUSIPCO,
  pages={2332--2336},
  year={2024},
  organization={IEEE}
}

@article{lim2020hodge,
  title={Hodge Laplacians on graphs},
  author={Lim, L. H.},
  journal={Siam Review},
  volume={62},
  number={3},
  pages={685--715},
  year={2020},
  publisher={SIAM}
}

@inproceedings{hoppe2024representing,
  title={Representing edge flows on graphs via sparse cell complexes},
  author={Hoppe, J. and Schaub, M. T.},
  booktitle={Learning on Graphs Conference},
  year={2024},
  organization={PMLR}
}

@inproceedings{isufi2022convolutional,
  title={Convolutional filtering in simplicial complexes},
  author={Isufi, E. and Yang, M.},
  booktitle=ICASSP,
  pages={5578--5582},
  year={2022},
  organization={IEEE}
}

@inproceedings{sardellitti2023probabilistic,
  title={Probabilistic topological models over simplicial complexes},
  author={Sardellitti, S. and Barbarossa, S.},
  booktitle=ASILOMAR,
  pages={822--826},
  year={2023},
  organization={IEEE}
}

@article{ghrist2008barcodes,
  title={Barcodes: the persistent topology of data},
  author={Ghrist, R.},
  journal={Bulletin of the American Mathematical Society},
  volume={45},
  number={1},
  pages={61--75},
  year={2008}
}

@article{edelsbrunner2003shape,
  title={On the shape of a set of points in the plane},
  author={Edelsbrunner, H. and Kirkpatrick, D. and Seidel, R.},
  journal=TIT,
  volume={29},
  number={4},
  pages={551--559},
  year={2003},
  publisher={IEEE}
}

@inproceedings{wang2019heterogeneous,
  title={Heterogeneous graph attention network},
  author={Wang, X. and Ji, H. and Shi, C. and Wang, B. and Ye, Y. and Cui, P. and Yu, P. S.},
  booktitle={The world wide web conference},
  pages={2022--2032},
  year={2019}
}

@ARTICLE{wu2024sclearning,
  author={Wu, H. and Yip, A. and Long, J. and Zhang, J. and Ng, M. K.},
  journal={IEEE Trans. Pattern Anal. Mach. Intell.}, 
  title={Simplicial Complex Neural Networks}, 
  year={2024},
  volume={46},
  number={1},
  pages={561-575}
  }

@inproceedings{gurugubelli2024sann,
  title={{SaNN}: Simple yet powerful simplicial-aware neural networks},
  author={Gurugubelli, S. and Chepuri, S. P.},
  booktitle=ICLR,
  year={2024}
}

@article{yang2025hodge,
  title={Hodge-Aware Convolutional Learning on Simplicial Complexes},
  author={Yang, M. and Leus, G. and Isufi, E.},
  journal={Trans. Mach. Learn. Res.},
  year={2025}
}

@article{giannakis2018topology,
  title={Topology identification and learning over graphs: Accounting for nonlinearities and dynamics},
  author={Giannakis, Georgios B and Shen, Yanning and Karanikolas, Georgios Vasileios},
  journal={Proceedings of the IEEE},
  volume={106},
  number={5},
  pages={787--807},
  year={2018},
  publisher={IEEE}
}

@inproceedings{marinucci2026simplicial,
  title={Simplicial Gaussian models: Representation and inference},
  author={Marinucci, L. and D’Acunto, G. and Di Lorenzo, P. and Barbarossa, S.},
  booktitle=ICASSP,
  pages={116--120},
  year={2026},
  organization={IEEE}
}
\end{document}